\documentclass[10pt,a4paper]{article}

\usepackage{a4wide}
\usepackage{mathtools}
\usepackage{amsfonts}
\usepackage{graphicx}
\usepackage{dcolumn}
\usepackage{bm}
\usepackage{url}
\usepackage{enumerate}
\usepackage{setspace}
\usepackage[dvipsnames]{xcolor}
\usepackage[sort&compress,comma,numbers,square]{natbib}

\begin{document}

\textbf{\Large{Predicting permeability via statistical learning on higher-order microstructural information}} \\

Magnus R\"oding$^{1,\star}$, Zheng Ma$^{2}$, Salvatore Torquato$^{3}$

$^{1}$ RISE Research Institutes of Sweden, G\"oteborg 41276, Sweden \\
$^{2}$ Department of Physics, Princeton University, Princeton, New Jersey 08544, USA \\
$^{3}$ Department of Chemistry, Department of Physics, Princeton Institute for the Science and Technology of Materials, and Program in Applied and Computational Mathematics, Princeton University, Princeton, New Jersey 08544, USA \\

$^{\star}$ Corresponding author: \url{magnus.roding@ri.se} \\
%$^{+}$ These authors contributed equally to this work \\

\section*{Abstract}
Quantitative structure-property relationships are crucial for the understanding and prediction of the physical properties of complex materials. For fluid flow in porous materials, characterizing the geometry of the pore microstructure facilitates prediction of permeability, a key property that has been extensively studied in material science, geophysics and chemical engineering. In this work, we study the predictability of different structural descriptors via both linear regressions and neural networks. A large data set of 30,000 virtual, porous microstructures of different types, including both granular and continuous solid phases, is created for this end. We compute permeabilities of these structures using the lattice Boltzmann method, and characterize the pore space geometry using one-point correlation functions (porosity, specific surface), two-point surface-surface, surface-void, and void-void correlation functions, as well as the geodesic tortuosity as an implicit descriptor. Then, we study the prediction of the permeability using different combinations of these descriptors. We obtain significant improvements of performance when compared to a Kozeny-Carman regression with only lowest-order descriptors (porosity and specific surface). We find that combining all three two-point correlation functions and tortuosity provides the best prediction of permeability, with the void-void correlation function being the most informative individual descriptor. Moreover, the combination of porosity, specific surface, and geodesic tortuosity provides very good predictive performance. This shows that higher-order correlation functions are extremely useful for forming a general model for predicting physical properties of complex materials. Additionally, our results suggest that artificial neural networks are superior to the more conventional regression methods for establishing quantitative structure-property relationships. We make the data and code used publicly available to facilitate further development of permeability prediction methods.

\section*{Introduction}
The study of how the microstructural morphology of random, heterogeneous, porous materials affects their effective properties, i.e., determining quantitative structure-property relationships, is key for the understanding and prediction of the physical properties of complex materials \cite{Torquato2013}. Specifically, understanding how fluid transport properties are related to the microstructure of a porous medium is crucial in a wide range of areas e.g. geological events \cite{vasseur2020permeability}, polymeric composites for packaging materials \cite{Silvestre2011}, catalysis, filtration and separation \cite{Slater2015}, energy, fuels, and electrochemistry \cite{Stamenkovic2017}, fiber and textile materials for health care and hygiene \cite{Langenhove2007}, and porous, biodegradable polymer films for controlled release of medical compounds \cite{Marucci2013}. Numerous efforts in determining the physical properties of complex materials have been made since the early work of Maxwell \cite{Torquato2013, milton2003theory, sahimi2011flow}, and such investigations have been enhanced due to the availability of high-resolution 3D images of various types of materials microstructures using X-ray nanotomography \cite{blunt2013pore, Lee2017} or focused ion beam scanning electron microscopy \cite{Gunda2011}.

The porosity $\phi$ (volume fraction of the pore phase) and specific surface $s$ (pore-solid interface area per unit volume) are perhaps the most basic geometrical characteristics. These two characteristics are the most frequently used in empirical expressions for the permeability. The Kozeny-Carman equation \cite{Kozeny1927, Carman1937} is the most notable example, usually written as
\begin{equation} \label{eq:kozeny}
k=\frac{\phi^3}{cs^2},
\end{equation}
where $k$ is the permeability and $c$ is the Kozeny-Carman constant. However, the remarkably simple form comes with great limitations. The Kozeny-Carman constant was found not to be a universal quantity. It does not only vary for different systems, but can also depend on the porosity \cite{kaviany2012principles}. Additionally, it does not distinguish portions of pore space that carries significant flow from portions that do not \cite{Torquato2013}.

To tackle this difficulty, countless modified versions of the original Kozeny-Carman equation have been proposed. However, these models are usually \textit{ad hoc} and only applicable to a specific class of structures \cite{Xu2008}. More importantly, although in many cases tortuosity is incorporated in the Kozeny-Carman constant \cite{mauret1997transport, mota2001binary, du1991flow, ahmadi2011analytical}, it usually only depends on the porosity alone in simplified models, thus the final expression of the permeability is essentially nothing more than a function of porosity and specific surface, i.e., $f(\phi)/s^2$. However, it is well-known that the microstructure is highly degenerate given only porosity and specific surface \cite{jiao2009superior, gommes2012microstructural}, which leads to a wide range of permeabilities as we show later. Thus, any function of the form $f(\phi)/s^2$ cannot be a general predictor and suffers from the intrinsic variances in the set of infinite degenerate microstructures.

Indeed, accurate prediction of the effective physical properties of the porous media requires a complete quantitative characterization of the microstructure in $d$-dimensional Euclidean space $\mathbb{R}^d$ via a variety of $n$-point correlation functions \cite{Torquato2013}. However, while such complete structural information about the medium is generally not available, reduced information in the form of lower-order correlation functions is often very beneficial. Two-point void-void and three-point void-void-void correlation functions have been used to produce both bounds and estimates for the effective electrical conductivity, diffusion coefficient and permeability \cite{Torquato1991, Jiao2012, Prager1961, Weissberg1962, Weissberg1970, Berryman1985a, Berryman1985b, Rubinstein1989, Liasneuski2014, Hlushkou2015}. In addition to the void-void correlation function, two-point surface-surface and surface-void correlation functions (where the surface is the interfacial surface between two phases) can also be defined and provide improved reconstructions of two-phase media from imaging data \cite{Zachary2011, Guo2014}, as well as sharper bounds on permeability compared to only using the void-void correlation function \cite{Torquato2013}.

On the other hand, it has been shown that permeability can be simply connected to the electrical formation factor of the porous material \cite{katz1986quantitative, torquato2020predicting}. This has been proved rigorously by Avellaneda and Torquato \cite{avellaneda1991rigorous}. However, from a prediction point of view, the formation factor itself needs to be measured experimentally or solved numerically, thus is not that helpful for establishing an explicit link to the microstructure. Although the formation factor is related to the hydraulic tortuosity \cite{Ghanbarian2013}, the later also requires heavy computations.

As a complement to rigorous approaches to estimate effective properties from the microstructure, data-driven methodologies to establish structure-property relationships are increasingly being used \cite{Linden2016, Stenzel2016, Neumann2019, Barman2019, Kondo2017, Wu2018, Sudakov2019}. The rapid increase in computational resources facilitates the computation of effective properties for very large (hundreds or thousands) of different microstructures. Moreover, as noted above, affordable high-resolution 3D digitized images of actual microstructures provide valuable data sets. As a consequence, it becomes manageable to generate large numbers of realistic virtual microstructures, and using those to perform exploratory computational screening of structure-property relationships. For example, van der Linden \emph{et al} \cite{Linden2016} use a data set of 536 virtual granular materials, compute 27 geometrical descriptors and use log-linear regression and other statistical learning methods as well as different variable selection schemes to understand the usefulness of the different descriptors for predicting permeability in these systems. Stenzel \emph{et al} \cite{Stenzel2016} study effective conductivity prediction in 43 virtual realizations of a stochastic spatial network model structure, using porosity and different tortuosity and constrictivity measures. This study was extended to 8,119 microstructures \cite{Stenzel2017}, which is likely the largest study published before, and the same data set was used again later to predict effective conductivity and permeability \cite{Neumann2019}. Barman \emph{et al} \cite{Barman2019} studied effective diffusivity prediction in 36 virtual porous polymer films using tortuosity and constrictivity. In a different direction, there are several attempts to use 2D and 3D convolutional neural networks (CNNs) to extract information directly from the binary image data describing the structure \cite{Kondo2017, Wu2018, Sudakov2019, Kamrava2020, Wu2019, Lubbers2017} in order to predict effective properties. However, these models are usually difficult to interpret and hard to rescale.

In this work, we are primarily interested in the predictive power of the information content contained in different microstructural descriptors. Specifically, we investigate the two-point surface-surface, surface-void, and void-void correlation functions, and also porosity, specific surface, and geodesic tortuosity using different regression methods. Unlike the hydraulic tortuosity mentioned above, the geodesic tortuosity is a purely geometric quantity that can be computed efficiently, and has been shown to be superior for diffusivity prediction \cite{Barman2019}. We compare different regression methods, including conventional linear regression with linear and quadratic terms, as well as deep artificial neural networks (deep learning). While conventional linear regression has an advantage in so far as the transparency of the prediction mechanism, deep learning has the potential to extract nearly the full information content of the descriptors, providing insight into the utility of the different descriptors for establishing the structure-property relationship. We find that the information content contained in these two-point correlation functions and geodesic tortuosity are indeed helpful to overcome the difficulty of applying a unique Kozeny-Carman-type equation to a variety of distinct microstructures, by yielding much better prediction performance. Moreover, our results suggest that artificial neural networks are superior to the more conventional regression methods for establishing quantitative structure-property relationships.

Consistent with the purpose of the paper, we have generated a large data set of virtual, porous, isotropic, and stationary microstructures of three different types, based on (i) thresholded Gaussian random fields, (ii) thresholded spinodal decomposition simulations of phase separation, and (iii) non-overlapping ellipsoid systems. Varying porosity and length scale and other parameters, we generate 10,000 structures of each of the three types, yielding a data set of 30,000 virtual microstructures in total. This is likely to be the largest data set of virtual microstructures ever created for studying permeability prediction, and covers both granular (ellipsoids) and continuous solid phases to provide a broad variability in the pore space geometry. Fluid flow is simulated using the lattice Boltzmann method. The large number of simulated microstructures makes it feasible to use not only scalar descriptors but also high-dimensional descriptors such as the two-point correlation functions, while still avoiding the well-known 'curse of dimensionality' in regression caused by having too many dimensions but too little data. To facilitate further investigation and development of permeability prediction methods, we make the microstructural descriptors, the computed permeabilities, the trained models, and the code used herein publicly available \cite{Zenodo2020}.

The paper is organized as follows. First, we introduce necessary definitions for the geometric descriptors used throughout the paper. Second, we describe how the virtual microstructures are generated, and the flow simulations and computations of permeability are described. Third, computation of the different microstructural descriptors is covered. Fourth, the prediction models for permeability are investigated. Finally, we make concluding remarks and discussions.

\section*{Background and definitions}
\subsection*{Geodesic Tortuosity}
We compute geodesic tortuosity in the flow direction according to Barman et al \cite{Barman2019} in the following manner. As a first step, a pointwise geodesic tortuosity is computed as $\tau\left(\mathbf{x}\right) = d\left(\mathbf{x}\right) / d$. Here, $d$ is the length of the microstructure in the flow direction, and $d\left(\mathbf{x}\right)$ is the length of the shortest path from any inlet pore to any outlet pore through $\mathbf{x}$. The shortest path is calculated as the sum of two geodesic distance transforms computed in the pore space of the binary voxel array: one using the set of edge voxels constituting the inlet pores as seeding points, and the other using the set of edge voxels constituting the outlet pores as seeding points. Let $\mathbb{P}$ be the set of voxels for which both geodesic distances are finite, i.e., the set of pore voxels connected to both inlet and outlet. Then, the geodesic tortuosity $\tau$ can be computed as
\begin{equation}
\tau = \left(\frac{1}{|\mathbb{P}|} \int_{\mathbf{x} \in \mathbb{P}} \frac{d\mathbf{x}}{\tau^2\left(\mathbf{x}\right)}\right)^{-1/2}.
\end{equation}
In Barman et al \cite{Barman2019}, it was found that accounting for both inlet and outlet in this manner is superior (in terms of diffusivity prediction) to just accounting for the inlet as is commonly done \cite{Stenzel2016, Pecho2015}. Tortuosity calculations were implemented in Matlab (Mathworks, Natick, MA, US).
\subsection*{Correlation functions}
Let $\mathcal{I}\left(\mathbf{x}\right)$ be the indicator function for the void phase (pore space) $\mathcal{V}_1$, i.e.,
\begin{equation}
\mathcal{I}\left(\mathbf{x}\right) =
\begin{cases}
    1, & \text{if } \mathbf{x} \in \mathcal{V}_1, \\
    0, & \text{otherwise}.
\end{cases}
\end{equation}
The two-point void-void correlation function is then generally defined by
\begin{equation}
S_2\left(\mathbf{x}_1, \mathbf{x}_2\right) = \langle \mathcal{I}\left(\mathbf{x}_1\right) \mathcal{I}\left(\mathbf{x}_2\right) \rangle.
\end{equation}
For statistically homogeneous materials, $S_2$ is only dependent on the vector difference $\mathbf{r} = \mathbf{x}_2 - \mathbf{x}_1$. Further, if the material is also statistically isotropic, $S_2$ is only dependent on the radial distance $r = |\mathbf{r}|$. Introducing a notation that is consistent with the other correlation functions defined below, the two-point void-void correlation function is now defined as
\begin{equation}
F_\mathrm{vv}\left(r\right) = \langle \mathcal{I}\left(\mathbf{x}\right) \mathcal{I}\left(\mathbf{x} + \mathbf{r}\right) \rangle,
\end{equation}
where the average is taken over all $\mathbf{x}$ and over all $\mathbf{r}$ with magnitude $r$. We proceed to the correlation functions involving the interfacial surface. Let $\mathcal{M}\left(\mathbf{x}\right)$ be the interface indicator function defined by \cite{Torquato2013}
\begin{equation}
\mathcal{M}\left(\mathbf{x}\right) = |\nabla \mathcal{I}\left(\mathbf{x}\right)|.
\end{equation}
Still assuming ergodicity and statistical isotropy, the surface-void correlation function can be written as
\begin{equation}
F_\mathrm{sv}\left(r\right) = \langle \mathcal{M}\left(\mathbf{x}\right) \mathcal{I}\left(\mathbf{x} + \mathbf{r}\right) \rangle,
\end{equation}
and the surface-surface correlation function can be written as
\begin{equation}
F_\mathrm{ss}\left(r\right) = \langle \mathcal{M}\left(\mathbf{x}\right) \mathcal{M}\left(\mathbf{x} + \mathbf{r}\right) \rangle.
\end{equation}

Importantly, the information content of one-point correlation functions (porosity $\phi$ and specific surface $s$) is automatically encoded into these two-point correlation functions. When $r$ goes to infinity, $F_\mathrm{vv}\left(r\right)$, $F_\mathrm{sv}\left(r\right)$ and $F_\mathrm{ss}\left(r\right)$ converge to $\phi^2$, $s\phi$ and $s^2$ respectively. Interestingly, the slope of $F_\mathrm{sv}\left(r\right)$ at the origin is proportional to the integrated mean curvature of the system \cite{Ma2018}, which has recently been shown to be a useful predictor of both permeabilities \cite{scholz2015direct} and diffusion coefficients \cite{howard2020connecting}.

Accurate and robust computation of $F_\mathrm{vv}$, $F_\mathrm{sv}$, and $F_\mathrm{ss}$ from discretized structures is a non-trivial task (in particular for the later two). The calculations recently became accessible due to the algorithms devised by Ma and Torquato \cite{Ma2018}. There, the calculations involving the interfacial surface are performed using a scalar field which when thresholded yields the corresponding two-phase medium. The details of the algorithms and the software can be found in Ref \cite{Ma2018}.
\begin{figure*}[]
\centering\includegraphics[width=150mm]{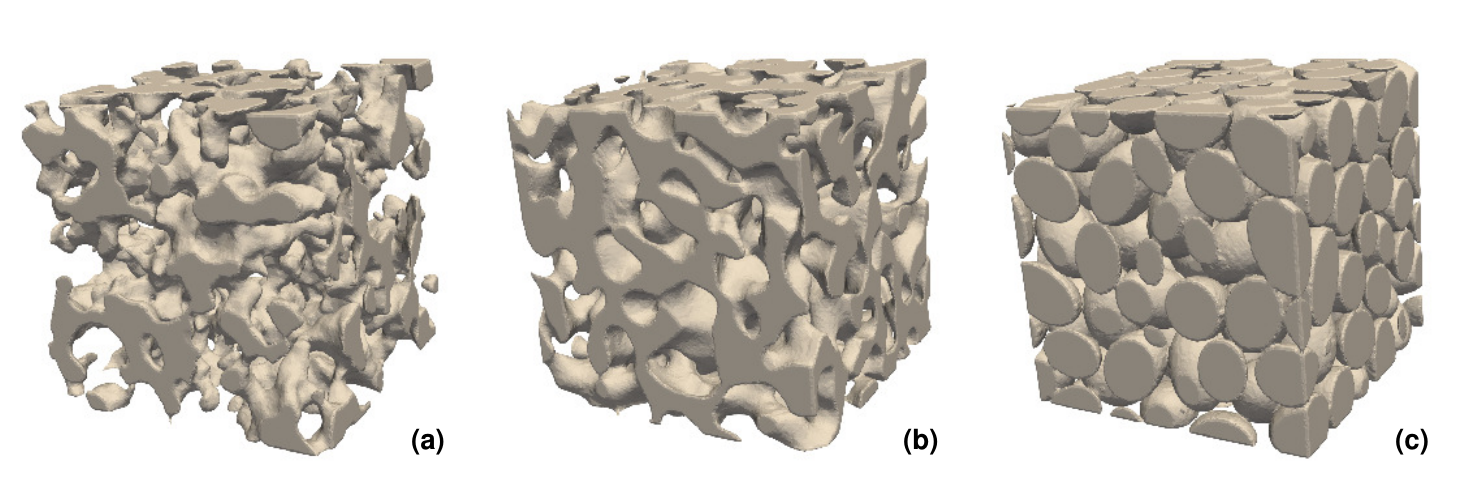}
\caption{\label{fig:structures} Examples of structures, showing (a) Gaussian random field structure with $\phi = 0.7$, (b) spinodal decomposition structure with $\phi = 0.5$, and (c) non-overlapping ellipsoid structure with $\phi = 0.3$.}
\end{figure*}

\section*{Microstructure data preparation}
\subsection*{Microstructure generation}
\begin{figure*}[]
\includegraphics[width=150mm]{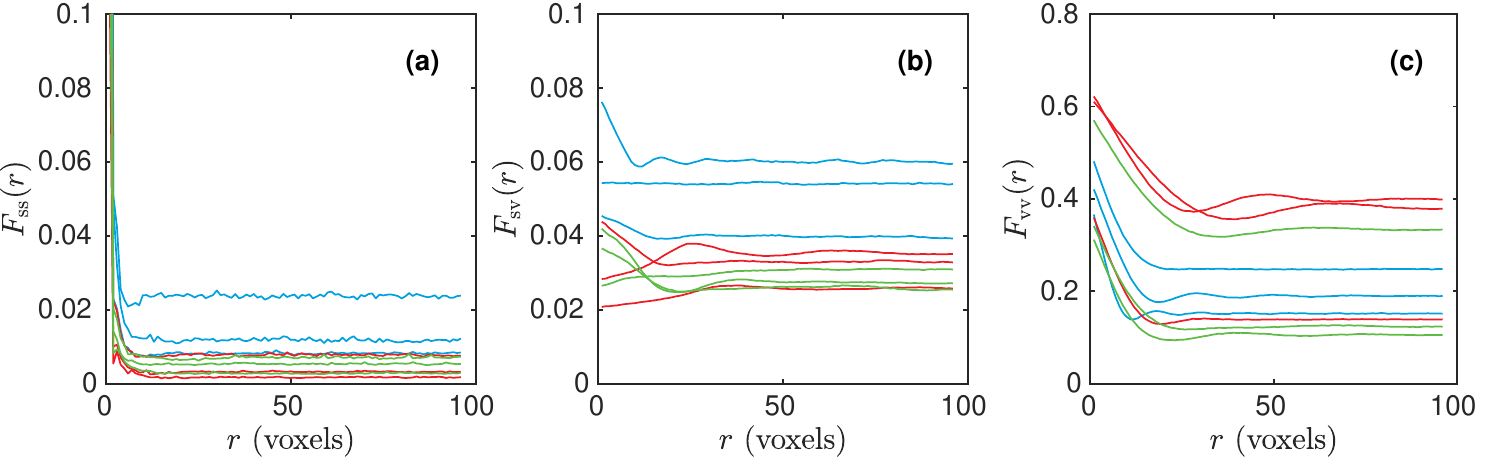}
\caption{\label{fig:correlation_functions} Some examples of (a) $F_\mathrm{ss}$, (b) $F_\mathrm{sv}$ , and (c) $F_\mathrm{vv}$ correlation functions. The examples are taken from the three different types of generated microstructures, i.e., thresholded Gaussian random fields (blue), thresholded spinodal decomposition simulations of phase separation (red), and non-overlapping ellipsoid systems (green).}
\end{figure*}

To achieve a large, representative data set, three different types of microstructures that are commonly studied in the materials literature are generated, including (i) thresholded Gaussian random fields, (ii) thresholded spinodal decomposition simulations of phase separation, and (iii) non-overlapping (hard) ellipsoid systems. We simulate $10,000$ realizations for each type, with porosities $\phi$ selected uniformly in $0.3 \le \phi \le 0.7$ and varying characteristic length scales. In the end, all structures are converted to $N^3$ binary voxel arrays with $N = 192$ voxels. In Fig.\ \ref{fig:structures}, one example of each type of structure is shown. We verified that the choice of the system volume size is both computationally manageable and representative. The correlation functions are evaluated for integer radii value bins $r$ from 1 to 96 voxels. In Fig.\ \ref{fig:correlation_functions}, some examples of correlation functions are shown. Note that these correlation functions are considerably distinct from each other, as seen by their different magnitudes and functional shapes. On the other hand, they have already converged to the large-$r$ limits within the sample size. The details of how these samples are generated are presented in the following subsections.

\subsubsection*{Gaussian random fields}
Gaussian random fields are generated according to Lang and Potthoff \cite{Lang2011}. Assuming that we wish to simulate a Gaussian random field $\mathcal{G}\left(\mathbf{x}\right)$, $\mathbf{x} \in \mathbb{R}^3$, with mean zero and covariance function $\Psi\left(\mathbf{x}, \mathbf{y}\right)$, it utilizes the fact that the covariance function can be written
\begin{equation}
\Psi\left(\mathbf{x}, \mathbf{y}\right) = \int_{\mathbb{R}^3} e^{-2 \pi i \left\langle \mathbf{p}, \mathbf{x}-\mathbf{y} \right\rangle } \gamma \left(\mathbf{p}\right) d\mathbf{p},
\end{equation}
where $\gamma \left(\mathbf{p}\right)$ is the spectral density of the Gaussian random field and $\left\langle \cdot, \cdot \right\rangle$ is the inner product. We wish to generate structures with length scale parameter $L$ and resolution $N^3$ voxels. Letting $\delta = L / N$ and letting FFT and FFT$^{-1}$ denote the forward and inverse 3-dimensional Fast Fourier Transforms, this can be performed in the following fashion: Generate an array $W$ where all elements are independent and normal distributed with mean zero and standard deviation $\delta^{-3}$ (white noise). Compute $\mathrm{FFT}(W)$. Define the Fourier space grid by $\mathbf{p} = \left(p_1, p_2, p_3\right)$, where $p_1 \in \{- N/(2 L), (- N/2 + 1)/L, ..., (N/2 - 2)/L, (N/2 - 1)/L\}$ and likewise for $p_2$ and $p_3$. Compute $\gamma \left(\mathbf{p}\right)$ on the grid. Compute $U = \mathrm{FFT}(W) \left(\mathbf{p}\right) \times \gamma \left(\mathbf{p}\right)^{1/2} / L^3$. Then, obtain the Gaussian random field as FFT$^{-1}(U)$.
We use several different spectral densities. For type (I),
\begin{equation}
\gamma\left(\mathbf{p}\right) = \left[ 1 + \left( p_1^2 + p_2^2 + p_3^2 \right)^l \right]^{-n}
\end{equation}
for $n = 1.95$ and $l = 1.85$ (power-law) \cite{Matern1986, Lang2011}. For type (II),
\begin{equation}
\gamma\left(\mathbf{p}\right) = \exp{\left[-\alpha^2 \left( p_1^2 + p_2^2 + p_3^2 \right)^{1/2} \right]},
\end{equation}
for $\alpha = 1.75$ (exponential). For type (III),
\begin{equation}
\gamma\left(\mathbf{p}\right) = \exp{\left[-\alpha^2 \left( p_1^2 + p_2^2 + p_3^2 \right) \right]},
\end{equation}
for $\alpha = 1.25$ (Gaussian). For type (IV),
\begin{equation}
\gamma\left(\mathbf{p}\right) =
\begin{cases}
    1, & \text{if } p_1^2 + p_2^2 + p_3^2 \le \rho^2, \\
    0, & \text{otherwise},
\end{cases}
\end{equation}
where $\rho = 1.25$ (circular top-hat). The parameters are chosen such that the corresponding Gaussian random fields have approximately the same characteristic spatial scale. For each spectral density, 2,500 structures are generated using uniformly distributed values of $L$, $4 \le L \le 16$. Thresholding then converts the scalar fields to corresponding two-phase media. To obtain a microstructure with a prescribed porosity $\phi$, the threshold is chosen to be an appropriate percentile of the values of $\mathcal{G}$. The method is implemented in Matlab (Mathworks, Natick, MA, US).

\subsubsection*{Spinodal decomposition}
The lattice Boltzmann method \cite{Geback2014, Geback2015}, a numerical framework for solving partial differential equations based on kinetic theory, is used to simulate phase separation kinetics (spinodal decomposition) using the Navier-Stokes and Cahn-Hilliard equations. Very briefly, the time evolution of a spatially dependent concentration $C\left(\mathbf{x}, t\right)$, $0 \le C \le 1$, is described by
\begin{equation}
\frac{\partial C}{\partial t} + \mathbf{u} \cdot \nabla C = M \nabla^2 \mu.
\end{equation}
As an initial condition, the values of $C$ are uniformly distributed in $0 \le C \le 1$, independently in all grid points. The phase separation is the coarsening of regions with $C \approx 0$ and $C \approx 1$ (ideally equal to 0 and 1). Here, $\mathbf{u}$ is a fluid velocity governed by the Navier-Stokes equations, $M$ is a mobility, i.e., a diffusion coefficient, and $\mu$ is the chemical potential. The simulation is performed using a dimensionless time step unity and both the density ratio and viscosity ratio between the phases are unity. The interface width, i.e., the characteristic length scale of the transition between the phases is 5 voxels. The simulations are performed in the resolution $96^3$ voxels with periodic boundary conditions, and are run until an appropriate degree of coarsening is obtained. The number of iterations $K$ is chosen randomly between 5 and 20,000 such that $K^{1/3}$ is approximately uniformly distributed; this is because according to the Lifschitz-Slyozov law, the typical length scale in the structure will be proportional to the cubic root of the simulation time. After terminating the simulation, the solutions are upscaled to $192^3$ voxels and thresholded to obtain the desired porosity. The spinodal decomposition simulations are run using in-house software \cite{Geback2014, Geback2015}.

\subsubsection*{Non-overlapping ellipsoids}
Random configurations of non-overlapping, hard ellipsoids are generated using a hard particle Markov Chain Monte Carlo (MCMC) algorithm. The Perram-Wertheim criterion \cite{Perram1985} for two ellipsoids of arbitrary orientation is used for overlap detection. First, particles are assigned uniformly distributed locations and orientations (the latter encoded using a quaternion representation). Second, the configurations are relaxed by sequentially performing random translations of all particles and then random rotations of all particles until no two particles overlap. Proposed translations and rotations are only accepted if they lead to a lower or equal degree of overlap for the considered particle. These ``local" stochastic optimization steps eventually lead to a ``global" optimization resulting in no overlap. Third, the configurations are equilibrated by performing a large number of random translations and rotations, ensuring a distribution in location and orientation that is as uniform as possible. Now, if the desired porosity $\phi$ is larger than 0.50, non-overlapping configurations can be generated easily at constant porosity as described above. Otherwise, as a final step, the configuration is further compressed in small steps, $\Delta \phi = 10^{-5}$, until the target porosity $\phi_\mathrm{target}$ is reached (in some cases, the configuration becomes jammed before reaching $\phi_\mathrm{target}$). The proposed translations are normal distributed with standard deviation $\sigma_\mathrm{t}$ in each direction. The proposed rotations are normal distributed with standard deviation $\sigma_\mathrm{r}$ in a random direction. In every step, $\sigma_\mathrm{t}$ and $\sigma_\mathrm{r}$ are chosen in an adaptive fashion to aim for an acceptance probability of 0.25. The number of ellipsoids $M$ is distributed in $8 \le M \le 512$ such that $M^{1/3}$ is approximately uniformly distributed, yielding an approximately uniform distribution of length scales. Further, the ellipsoids have semi-axes $(1,1,\eta)$ where $\eta$ is uniform in $0.25 \le \eta \le 1$ (oblate) with probability 0.5 and otherwise uniform in $1 \le \eta \le 4$ (prolate). The random microstructures are generated using in-house developed software implemented in Julia (\url{www.julialang.org}) \cite{Bezanson2017} and available in a Github repository (\url{https://github.com/roding/whitefish_generation}, version 0.2). The obtained configurations are further smoothed with a Gaussian filter with $\sigma = 3$ voxels and thresholded again to regain the original porosity; the reason for this is that computation of some of the correlation functions requires the binary structures to be described as a thresholded version of smooth scalar fields.

\subsection*{Flow simulations}
\begin{figure}[h!]
\centering\includegraphics[width=80mm]{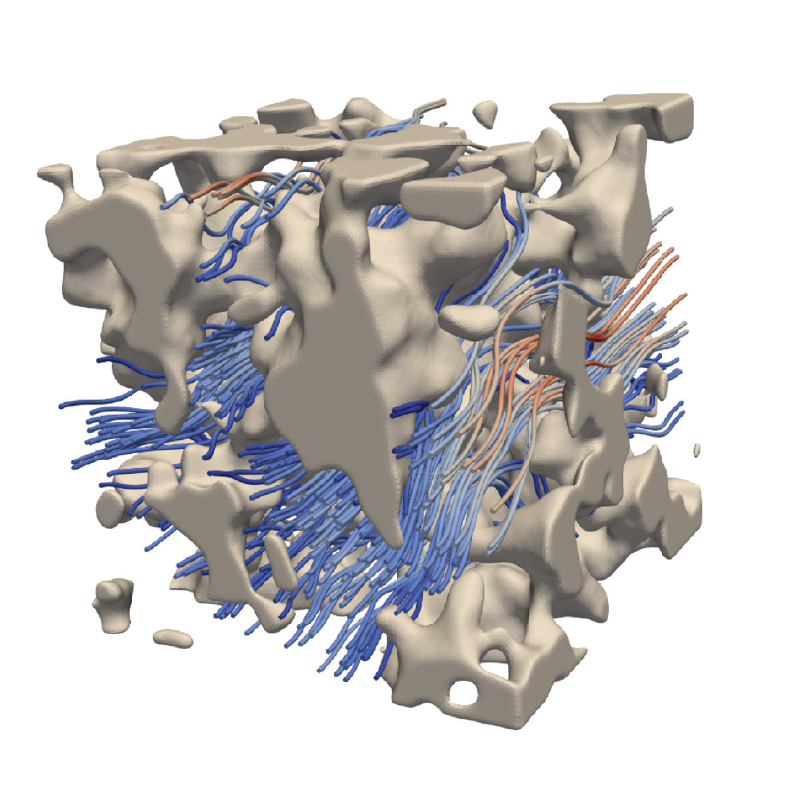}
\caption{\label{fig:structure_simulation} An example of a simulated steady state flow through a Gaussian random field microstructure with porosity $\phi = 0.7$. Regions with slow and fast flow are indicated by blue and red flow lines, respectively.}
\end{figure}

The lattice Boltzmann method \cite{Geback2014, Geback2015}, a numerical framework for solving partial differential equations based on kinetic theory, is used to simulate fluid flow through the structures. The Navier-Stokes equations for pressure-driven flow are solved for the steady state using no-slip, bounce-back boundary conditions on the solid/liquid interface and periodic boundary conditions orthogonal to the flow direction. We use the two relaxation time collision model with the free parameter $\lambda_{eo}=\frac{3}{16}$, which guarantees that the computed permeability is independent of the relaxation time (and thus the viscosity) \cite{Ginzburg2008}. The relaxation time $\tau = -\frac{1}{\lambda_e}$ is kept at $1.25$. The flow is driven by constant pressure difference boundary conditions across the structure in the primary flow direction \cite{Zou1997}, and a linear gradient is used as initial condition. The computational grid coincides with the voxels of the binary structure, i.e., it has $192^3$ grid points. After convergence to steady state flow, the permeability $k$ is obtained from Darcy's law,
\begin{equation}
\bar u = - \frac{k \Delta p}{\mu d}.
\label{eq:darcy}
\end{equation}
Here, $\bar u$ is the average velocity, $\Delta p$ is the applied pressure difference, $\mu$ is the dynamic viscosity, and $d$ is the length of the microstructure in the flow direction. The permeability is independent of the fluid and the pressure difference and a property solely of the microstructure provided that the Reynolds number is sufficiently small ($\mathrm{Re} < 0.01$), which also ensures that the velocity is proportional to the pressure difference. The computed permeabilities have units voxels$^2$, where the voxels have unit length. Fig.\ \ref{fig:structure_simulation} illustrates the result of a flow simulation in one of the Gaussian random field structures.

\begin{figure}[h!]
\centering\includegraphics[width=80mm]{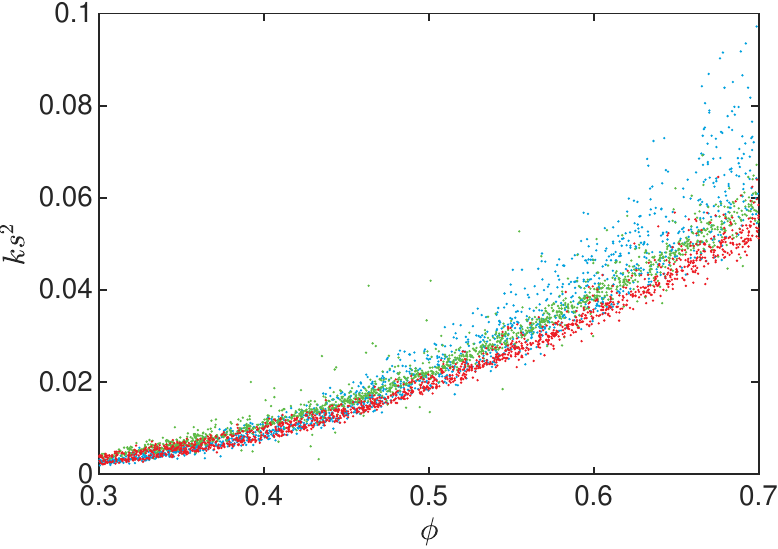}
\caption{\label{fig:phi_vs_ks2} A scatter plot of scaled permeabilities $ks^2$ versus porosities $\phi$ for 4,500 microstructures, 1,500 for each type, i.e., thresholded Gaussian random fields (blue), non-overlapping ellipsoid systems (green) and spinodal decomposition simulations (red).}
\end{figure}

We choose 4,500 miscrostructures, 1,500 for each type, and plot their scaled permeabilities $ks^2$ versus porosities $\phi$ in Fig.\ \ref{fig:phi_vs_ks2}. It is noteworthy that although a clear overall trend can be seen, the scaled permeability is never a function of $\phi$ alone. In fact, we observe that for the same porosity, the largest scaled permeability is approximately twice as large as the smallest one. This degeneracy of microstructures clearly show why the Kozeny-Carman equation and some of its modifications typically fail for general structures. Consequentially, more detailed information is needed to pinpoint the true permeability on this ``band".

Another interesting observation is that the scaled permeabilities of spinodal decomposition patterns are almost always lower than those of the other two types. It has been shown that spinodal decomposition gives rise to hyperuniform structures in the scaling region \cite{ma2017random}. This observation is consistent with the fact that hyperuniform structures cannot tolerate large ``holes" and the pore space is more evenly distributed compared to nonhyperuniform structures, thus their permeabilities are generally lower \cite{torquato2020predicting}.

\section*{Microstructural descriptors}
We study the performance of the different microstructural descriptors introduced above and the combinations of them for predicting the permeability $k$ (dimension length$^2$). The descriptors used are porosity $\phi$ (dimensionless), specific surface $s$ (dimension 1/length), tortuosity $\tau$ (dimensionless), the correlation functions $F_\mathrm{ss}$ (dimension 1/length$^2$), $F_\mathrm{sv}$ (dimension 1/length), and $F_\mathrm{vv}$ (dimensionless). Additionally, we investigate a particular combination of the correlation functions: inspired by a rigorous upper bound for permeability in isotropic media \cite{Ma2018}, i.e.,
\begin{equation}
k \le \frac{2}{3} \int_{0}^{\infty} \left[ \frac{\phi^2}{s^2} F_\mathrm{ss}(r) - \frac{2\phi}{s} F_\mathrm{sv}(r) + F_\mathrm{vv}(r) \right] r \mathrm{d}r,
\label{eq: sbound}
\end{equation}
we define a function
\begin{equation}
F(r) = \frac{\phi^2}{s^2} F_\mathrm{ss}(r) - \frac{2\phi}{s} F_\mathrm{sv}(r) + F_\mathrm{vv}(r),
\end{equation}
which is also used for prediction ($F$ is dimensionless and converges to zero when $r$ goes to infinity). Each correlation function is represented by a 96-dimensional vector. In Table \ref{tab:descriptors_k}, the models, denoted 1 through 7, and the sizes of the corresponding input features are listed.
\begin{table}[h!]
\centering
\begin{tabular}{p{0.75cm} p{4cm} p{2.5cm}}
\hline
No.                             & Descriptors                                                   & Input size    \\
\hline
1                               & $\phi$, $s$, $\tau$                                           & 3             \\
2                               & $F_\mathrm{ss}$                                               & 96            \\
3                               & $F_\mathrm{sv}$                                               & 96            \\
4                               & $F_\mathrm{vv}$                                               & 96            \\
5                               & $F$                                                           & 96            \\
6                               & $F_\mathrm{ss}$, $F_\mathrm{sv}$, $F_\mathrm{vv}$             & 288           \\
7                               & $F_\mathrm{ss}$, $F_\mathrm{sv}$, $F_\mathrm{vv}$, $\tau$     & 289           \\
\hline
\end{tabular}
\caption{Descriptors for prediction of the permeability $k$.}
\label{tab:descriptors_k}
\end{table}
Additionally, we consider a rescaling of the problem, predicting the rescaled, dimensionless permeability $k s^2$ instead of $k$ directly. For model 1, we remove $s$ from the descriptors since it is already absorbed in the permeability; for models 2, 3, 4, 6, and 7, the correlation functions are rescaled to dimensionless versions where applicable. In Table \ref{tab:descriptors_ks2}, the modified models, denoted 1$'$ through 7$'$, and the dimensions of the corresponding input vectors (that changes only for model 1$'$) are listed.\\
\begin{table}[h!]
\centering
\begin{tabular}{p{0.75cm} p{4cm} p{2.5cm}}
\hline
No.                             & Descriptors                                                       & Input size    \\
\hline
1$'$                            & $\phi$, $\tau$                                                    & 2             \\
2$'$                            & $F_\mathrm{ss}/s^2$                                               & 96            \\
3$'$                            & $F_\mathrm{sv}/s$                                                 & 96            \\
4$'$                            & $F_\mathrm{vv}$                                                   & 96            \\
5$'$                            & $F$                                                               & 96            \\
6$'$                            & $F_\mathrm{ss}/s^2$, $F_\mathrm{sv}/s$, $F_\mathrm{vv}$           & 288           \\
7$'$                            & $F_\mathrm{ss}/s^2$, $F_\mathrm{sv}/s$, $F_\mathrm{vv}$, $\tau$   & 289           \\
\hline
\end{tabular}
\caption{Descriptors for prediction of the rescaled, dimensionless permeability $k s^2$.}
\label{tab:descriptors_ks2}
\end{table}
In practice, we use the logarithm of permeabilities, i.e., $\log_{10} k$ and $\log_{10} \left(k s^2\right)$, instead of their original values. The reason for using logarithms of the permeabilities is that they span several orders of magnitude. By taking logarithms, the predictions are simplified, and guaranteed to be positive. We also use the logarithms of porosity, specific surface, and tortuosity since we know that models 1 and 1$'$ are naturally multiplicative in these Kozeny-Carman-like equations.
\section*{Predictive models}
We assess the predictive performance of the different descriptors/inputs using several regression methods, namely, linear regression with linear terms only or combined with quadratic terms, and deep artificial neural networks. The inputs are as described above, with no normalization (such as subtracting feature-wise means; our investigation suggested no improvement from normalization in this setting). For each microstructure class, the data are split randomly into training data (70 \%; 7,000 per class), validation data (15 \%; 1,500 per class), and test data (15 \%; 1,500 per class). In total, the training, validation, and test data sets hence consist of 21,000, 4,500, and 4,500 samples, respectively. The split is kept fixed across all inputs and all regression methods. Training data is used for the actual estimation of a functional relationship mapping input to output. Validation data is used for hyperparameter selection, i.e., finding optimal values for e.g.\ learning rates for ANNs (in the case of linear regressions, the validation data is not used because we do not have any hyperparameters to optimize). Test data is used for final assessment of the predictive performance. To quantify error/loss in prediction, we use several different measures. Let $k$ be the 'true' permeability, i.e., the value obtained from the lattice Boltzmann simulations, and let $\hat k$ be the predicted value. We use mean squared error (MSE) in the logarithmic scale,
\begin{equation}
\mathrm{MSE} = \left\langle \left( \log_{10} \hat k - \log_{10} k \right)^2 \right\rangle,
\end{equation}
root mean squared error (RMSE) which is just $\mathrm{RMSE} = \mathrm{MSE}^{1/2}$, and mean absolute percentage error (MAPE) in the linear scale, i.e.,
\begin{equation}
\mathrm{MAPE} = 100 \times \left\langle \left|\frac{\hat k - k}{k}\right| \right\rangle \%.
\end{equation}
Using MSE is the most practical and most common choice for model fitting. However, for final assessment of performance, the linear scale and MAPE is a more straightforward and intuitive choice.

\subsection*{Linear regression with linear terms}
First, we consider using linear regression with only linear terms (i.e., only the input descriptors to the power of unity are used). For models 1 and 1$'$, this becomes multiplicative regression in a Kozeny-Carman-like form, i.e.,
\begin{equation}
\log_{10} k = c_0 + a \log_{10}\phi + b \log_{10} s + c \log_{10} \tau
\end{equation}
and
\begin{equation}
\log_{10} \left(k s^2\right) = c_0 + a \log_{10}\phi + c \log_{10} \tau.
\end{equation}
It is well established that $a > 0 $, $b < 0 $, and $c < 0 $ in this setting (and due to the dimensions, $b = - 2$ would be preferable).

On the other hand, the rationale behind the linear regression model of correlation functions is inspired by the rigorous bounds involving correlation functions, such as equation (\ref{eq: sbound}). Since the integral in the bounds can be seen as the inner product between the correlation functions and another predetermined function, it is natural to assume that a functional regression on the correlation functions may yield a reasonable estimation of permeabilities. For correlation functions evaluated on discrete grids, the model essentially becomes a linear regression model. To give a couple of examples of the regressions on correlation functions, model 2 becomes
\begin{equation}
\log_{10} k = c_0 + \sum_{i} \alpha(r_i) F_\mathrm{ss}(r_i),
\end{equation}
model 6 becomes
\begin{equation}
\log_{10} k = c_0 + \sum_{i} \alpha(r_i) F_\mathrm{ss}(r_i) + \sum_{i} \beta(r_i) F_\mathrm{sv}(r_i) + \sum_{i} \gamma(r_i) F_\mathrm{vv}(r_i),
\end{equation}
and model 3$'$ becomes
\begin{equation}
\log_{10} \left(k s^2\right) = c_0 + \sum_{i} \alpha(r_i) F_\mathrm{sv}(r_i)/s.
\end{equation}
The rest of the models are formulated in an equivalent fashion. For the correlation function-based models, $\alpha$, $\beta$, and $\gamma$ are just vectors of coefficients but can also be thought of as discretized forms of continuous coefficient functions $\alpha(r)$, $\beta(r)$, and $\gamma(r)$. We use least squares fitting, finding the coefficients that minimize the training set MSE. We also include the reference Kozeny-Carman model in this category as a benchmark. Fitting is performed in Matlab (Mathworks, Natick, MA, US).
\begin{table}[h!]
\centering
\begin{tabular}{p{0.75cm} p{1.75cm} p{1.75cm} p{1.75cm} p{1.75cm}}
\hline
            & RMSE          &           &           & MAPE (\%)     \\
No.         & Train         & Val       & Test      & Test          \\
\hline
\multicolumn{5}{l}{\textit{Kozeny-Carman model}}                    \\
-           & 0.084         & 0.084     & 0.086     & 14.986        \\
\multicolumn{5}{l}{\textit{Unscaled models}}                        \\
1           & 0.046         & 0.046     & 0.047     & 8.291         \\
2           & 0.370         & 0.371     & 0.376     & 98.028        \\
3           & 0.163         & 0.163     & 0.167     & 29.719        \\
4           & 0.068         & 0.069     & 0.069     & 12.011        \\
5           & 0.233         & 0.238     & 0.243     & 51.563        \\
6           & 0.049         & 0.050     & 0.051     & 8.351         \\
7           & 0.033         & 0.034     & 0.034     & 5.536         \\
\multicolumn{5}{l}{\textit{Rescaled models}}                        \\
1$'$        & 0.052         & 0.051     & 0.052     & 8.653         \\
2$'$        & 0.360         & 0.361     & 0.365     & 96.907        \\
3$'$        & 0.078         & 0.078     & 0.081     & 14.363        \\
4$'$        & 0.059         & 0.058     & 0.060     & 10.315        \\
5$'$        & 0.165         & 0.164     & 0.168     & 33.823        \\
6$'$        & 0.043         & 0.044     & 0.045     & 7.488         \\
7$'$        & 0.029         & 0.030     & 0.031     & 5.063         \\
\hline
\end{tabular}
\caption{RMSE for the training, validation, and test sets and MAPE for the test set for the regression models with linear terms. We also include the reference Kozeny-Carman model in this category.}
\label{tab:results_reg_lin}
\end{table}

Specifically, the fitted Kozeny-Carman model writes as
\begin{equation}
k = \frac{\phi^3}{6.23 s^2}.
\end{equation}
For models 1 and 1$'$, the estimated relationships become
\begin{equation}
k = 0.29 \frac{\phi^{2.68}}{s^{1.88} \tau^{7.28}},
\end{equation}
and
\begin{equation}
k s^2 = 0.21 \frac{\phi^{2.77}}{\tau^{6.36}}.
\end{equation}
Interestingly, the estimated coefficients for models 1 and 1$'$ are quite similar, and they are also comparable to the Kozeny-Carman model. Specifically, even without being forced to have the dimensionally correct exponent -2 for $s$, as in the case of model 1$'$, the estimated exponent from the regression (-1.88) in model 1 is very close.

\begin{figure}[h]
\centering\includegraphics[width=80mm]{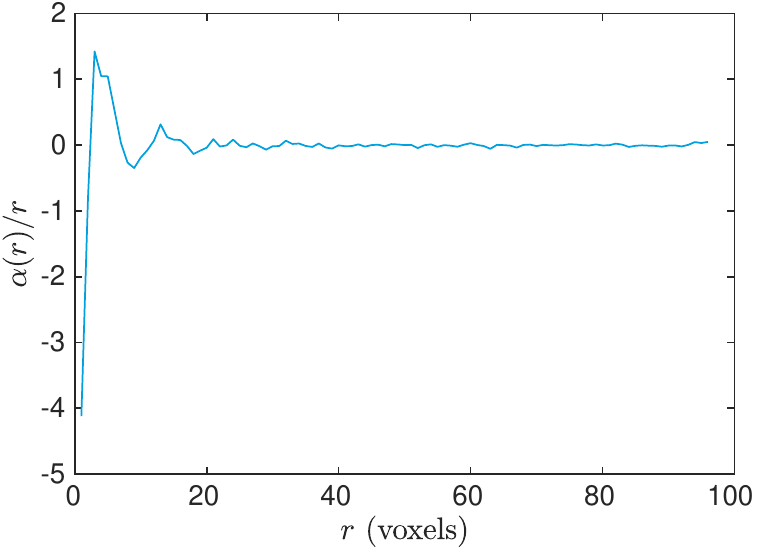}
\caption{\label{fig:figure_coeff_fun_model_4} The estimated coefficient function $\alpha(r)$ scaled by $r$ for model 4 using the linear regression model. One can see that it gives larger weight to the small-to-intermediate-$r$ behavior of $F_\mathrm{vv}$.}
\end{figure}

To get an idea of the dependence of the coefficient functions on $r$, we plot the estimated $\alpha(r)/r$ in Fig.\ \ref{fig:figure_coeff_fun_model_4} as an example. We scale $\alpha(r)$ by $r$ in order to make it have the appropriate weight consistent with the one that multiplies the correlation functions in the integrand of Eq. (\ref{eq: sbound}). It is clear that instead of weighting on different parts of $F_\mathrm{vv}(r)$ equally, the regression emphasizes on the small-to-intermediate-$r$ behavior of the correlation function as expected.

\begin{figure*}[]
\centering\includegraphics[width=165mm]{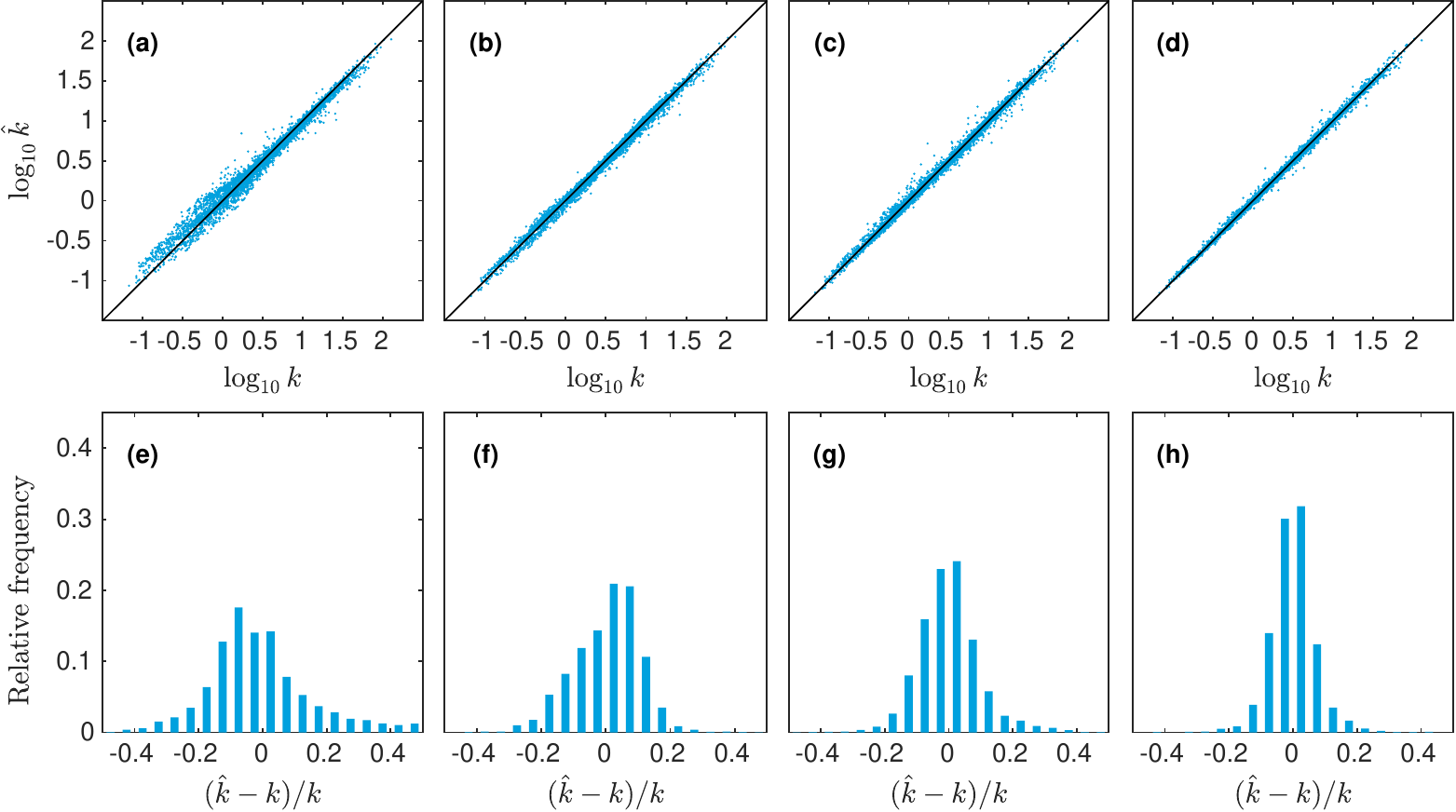}
\caption{\label{fig:figure_prediction_lin} Predicted values $\log_{10} \hat k$ vs the true (simulated) values $\log_{10} k$ for linear regression with linear terms, showing (a) the Kozeny-Carman model, (b) model 1, (c) model 6$'$, and (d) model 7$'$. In (e)-(h), histograms of relative errors are shown for the same set of models.}
\end{figure*}
We also made an attempt to further smooth the coefficient functions according to a proposed functional regression scheme published earlier \cite{Roding2017}. Briefly, the idea is to introduce a penalty term in the least squares objective function that is proportional to the integral of the squared second derivative of the coefficient function, such that the functional form becomes smoother. To exemplify, pick a model with a coefficient function $\alpha(r)$, and we minimize
\begin{equation}
\sum_{j} \left( \log_{10} \hat k(j) - \log_{10} k(j) \right)^2 + \lambda \sum_i \left(\alpha(r_{i-1}) - 2\alpha(r_{i}) + \alpha(r_{i+1})\right)^2
\end{equation}
for a range of penalty parameters $\lambda$, and pick $\lambda$ such that the validation MSE is minimized. Interestingly, the smoothing procedure provides only a negligible improvement in validation and test errors, and a negligible increase in smoothness of the coefficient functions. Thus, we stick to the models without penalties.

The errors for training, validation, and test sets for the aforementioned models are shown in Table \ref{tab:results_reg_lin}. Although we perform no hyperparameter optimization for these regression models (such as variable selection), we include the validation set errors for consistency. There are several noteworthy observations about our results. First, the Kozeny-Carman-like model (1 and 1$'$) with geodesic tortuosity almost reduce the relative error by half compared to the original Kozeny-Carman model, while the combination of all three correlation functions (6 and 6$'$) also give comparable performance. Interestingly, when we further combine the correlation functions with geodesic tortuosity (7 and 7$'$), the error shrinks to approximately only one-third of the error generated by the Kozeny-Carman model. Among single correlation functions, the best performance is given by $F_\mathrm{vv}$, while the worst is given by $F_\mathrm{ss}$. This result is expected, since $F_\mathrm{vv}$ alone can yield a bound on the permeability, while $F_\mathrm{ss}$ alone does not even contain the most important information, i.e., the porosity. The reason we keep model 2 and 2$'$ is indeed just for self-consistency. Interestingly, the compound correlation function $F$ performs relatively poorly. This is probably due to the fact that it washes out the information content contained in individual correlation functions. However, its error can be interpreted as a lower bound on how good the bound in equation (\ref{eq: sbound}) would work on our data set. To better visualize our findings, the predicted values vs the true (simulated) values and histograms of relative errors for a few selected models are shown in Fig.\ \ref{fig:figure_prediction_lin}. It is obvious that by adding geodesic tortuosity and correlation functions the prediction error can be greatly reduced. It can also be seen that although the Kozeny-Carman-like model (Fig.\ \ref{fig:figure_prediction_lin}b) and the correlation function based one (Fig.\ \ref{fig:figure_prediction_lin}c) has similar MAPE, the error distribution of the correlation function based one is more symmetric.

\subsection*{Linear regression with quadratic terms}
\begin{figure*}[]
\centering\includegraphics[width=165mm]{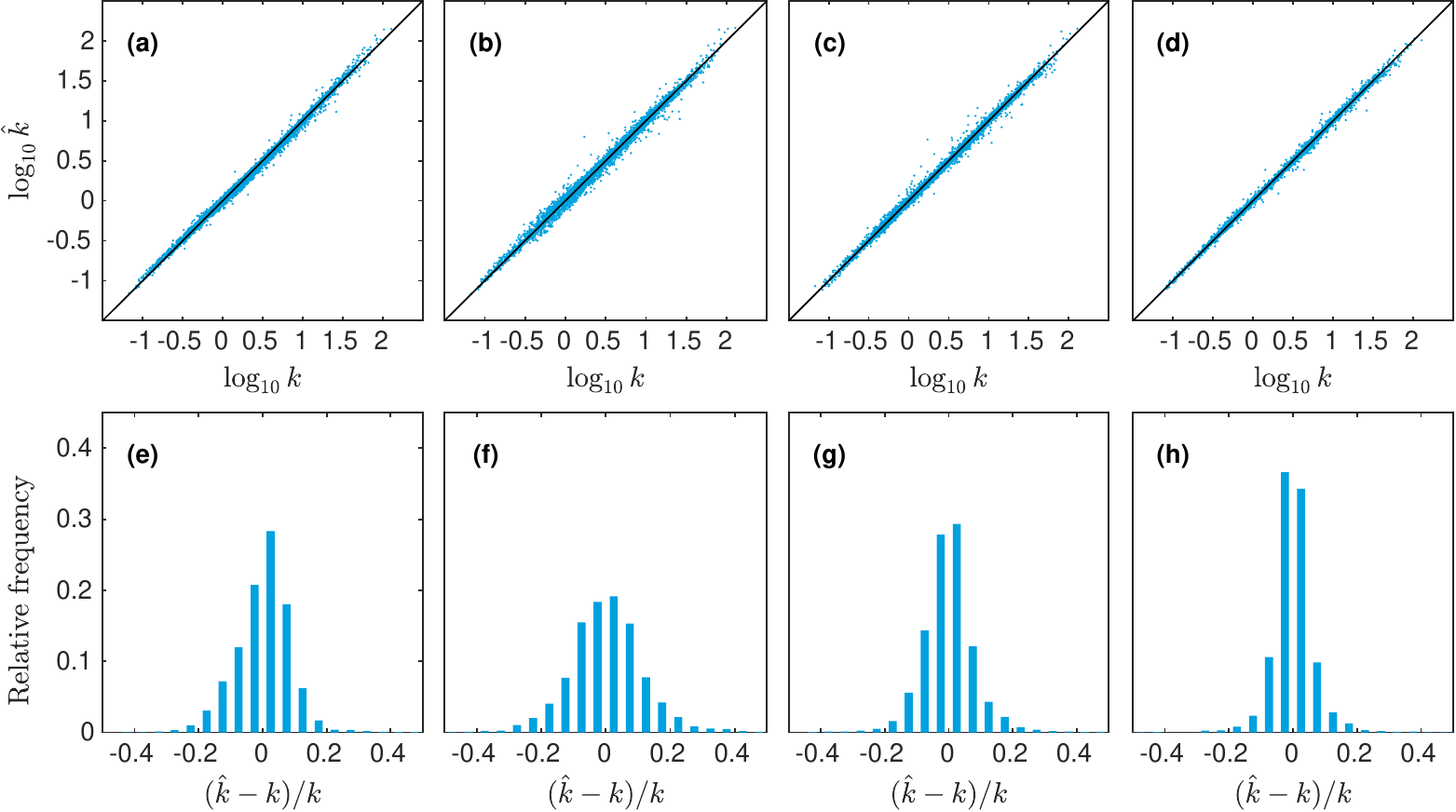}
\caption{\label{fig:figure_prediction_quad} Predicted values $\log_{10} \hat k$ vs the true (simulated) values $\log_{10} k$ for linear regression with linear and quadratic terms, showing (a) model 1, (b) model 4$'$, (c) model 6$'$, and (d) model 7. In (e)-(h), histograms of relative errors are shown for the same set of models.}
\end{figure*}

Second, we generalize the previous models that utilized linear terms only to incorporate both linear and quadratic terms. For the correlation function models, we use both the correlation functions themselves and their squares. For example, we use both $F_\mathrm{ss}$ and $F^2_\mathrm{ss}$ as input data in model 2. It is worth to point out that we use only pure quadratic terms, such as $F^2_\mathrm{ss}(r_i)$, but not mixed quadratic terms, such as $F_\mathrm{ss}(r_i) F_\mathrm{ss}(r_j)$ for $i \neq j$. We include models 1 and 1$'$ in this investigation as well, mainly for completeness, and add terms of the type $\left(\log_{10} \phi\right)^2$. Fitting is performed in Matlab (Mathworks, Natick, MA, US). The errors for training, validation, and test sets are again shown in Table \ref{tab:results_reg_quad}.
\begin{table}[h!]
\centering
\begin{tabular}{p{0.75cm} p{1.75cm} p{1.75cm} p{1.75cm} p{1.75cm}}
\hline
            & RMSE          &           &           & MAPE (\%)     \\
No.         & Train         & Val       & Test      & Test          \\
\hline
\multicolumn{5}{l}{\textit{Unscaled models}}                        \\
1           & 0.038         & 0.038     & 0.039     & 6.560         \\
2           & 0.351         & 0.351     & 0.359     & 92.349        \\
3           & 0.079         & 0.079     & 0.081     & 13.547        \\
4           & 0.059         & 0.061     & 0.061     & 10.624        \\
5           & 0.151         & 0.156     & 0.163     & 29.633        \\
6           & 0.039         & 0.040     & 0.040     & 6.377         \\
7           & 0.027         & 0.027     & 0.028     & 4.300         \\
\multicolumn{5}{l}{\textit{Rescaled models}}                        \\
1$'$        & 0.050         & 0.049     & 0.050     & 8.216         \\
2$'$        & 0.358         & 0.360     & 0.365     & 96.739        \\
3$'$        & 0.051         & 0.051     & 0.052     & 8.717         \\
4$'$        & 0.051         & 0.052     & 0.053     & 8.923         \\
5$'$        & 0.118         & 0.118     & 0.122     & 22.362        \\
6$'$        & 0.036         & 0.039     & 0.039     & 6.041         \\
7$'$        & 0.026         & 0.028     & 0.029     & 4.515         \\
\hline
\end{tabular}
\caption{RMSE for the training, validation, and test sets and MAPE for the test set for the regression models with linear and quadratic terms.}
\label{tab:results_reg_quad}
\end{table}
In Fig.\ \ref{fig:figure_prediction_quad}, the predicted values vs the true (simulated) values and histograms of relative errors for a few selected models are shown.

\begin{figure*}[]
\centering\includegraphics[width=165mm]{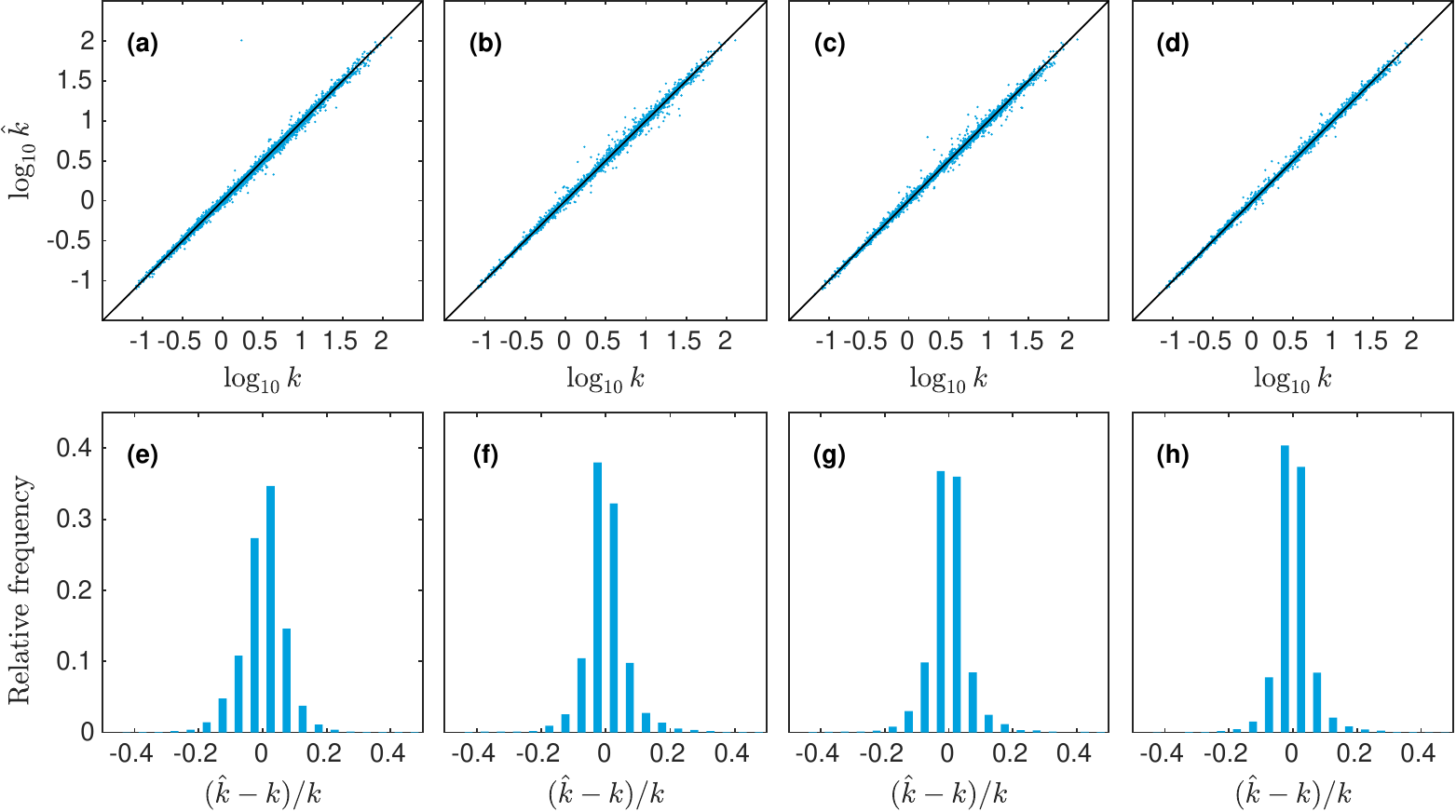}
\caption{\label{fig:figure_prediction_ann} Predicted values $\log_{10} \hat k$ vs the true (simulated) values $\log_{10} k$ for deep neural network regression, showing (a) model 1, (b) model 4, (c) model 6, and (d) model 7. In (e)-(h), histograms of relative errors are shown for the same set of models.}
\end{figure*}

Importantly, we note that adding the quadratic terms leads to an improvement for every model. This suggests that the relation between the microstructural descriptors and the permeability can be quite complex such that a simple linear model may not be able to fully capture it. However, the relative rank of performances roughly remain the same, showing the validity of our previous arguments. The estimated coefficient functions are quite noisy and their physical meaning is not obvious. We also make an attempt to use a full quadratic model, incorporating also mixed terms such as $F_\mathrm{ss}(r_i) F_\mathrm{ss}(r_j)$, or even mixed between correlation functions, such as $F_\mathrm{sv}(r_i) F_\mathrm{vv}(r_j)$. The numbers of variables in the models then become very large, leading to ill-conditioned estimation problems. We investigated whether the Lasso variable selection technique \cite{Tibshirani1996}, which forces a variable number of coefficients in a linear model to become zero by penalizing the sum of absolute values of the coefficients, could act as an efficient means of reducing the model dimensionality. However, it turns out that no amount of Lasso regularization can decrease the validation MSE in this case. There are two likely reasons for this: Lasso is primarily intended for high-dimensional variable spaces where a large fraction of the variables contain little information and mostly noise, and can be disregarded easily. This is likely not the case for the correlation functions. Also, because the values are taken from continuous functions, they are strongly correlated, which is known to compromise the underlying rationale of Lasso.

\subsection*{Neural networks}

The complexity of the linear regression models could be further increased, for example by incorporating pure cubic terms. Although we expect to see further improvements, the linear regression model can quickly become ill-conditioned and intractable on this track. For this reason, we proceed to consider deep neural networks, which can potentially fully capture the complex structure-property relationships. Thus we can exploit the complete information content contained in the descriptors.

\begin{figure}[]
\centering\includegraphics[width=86mm, height=55mm]{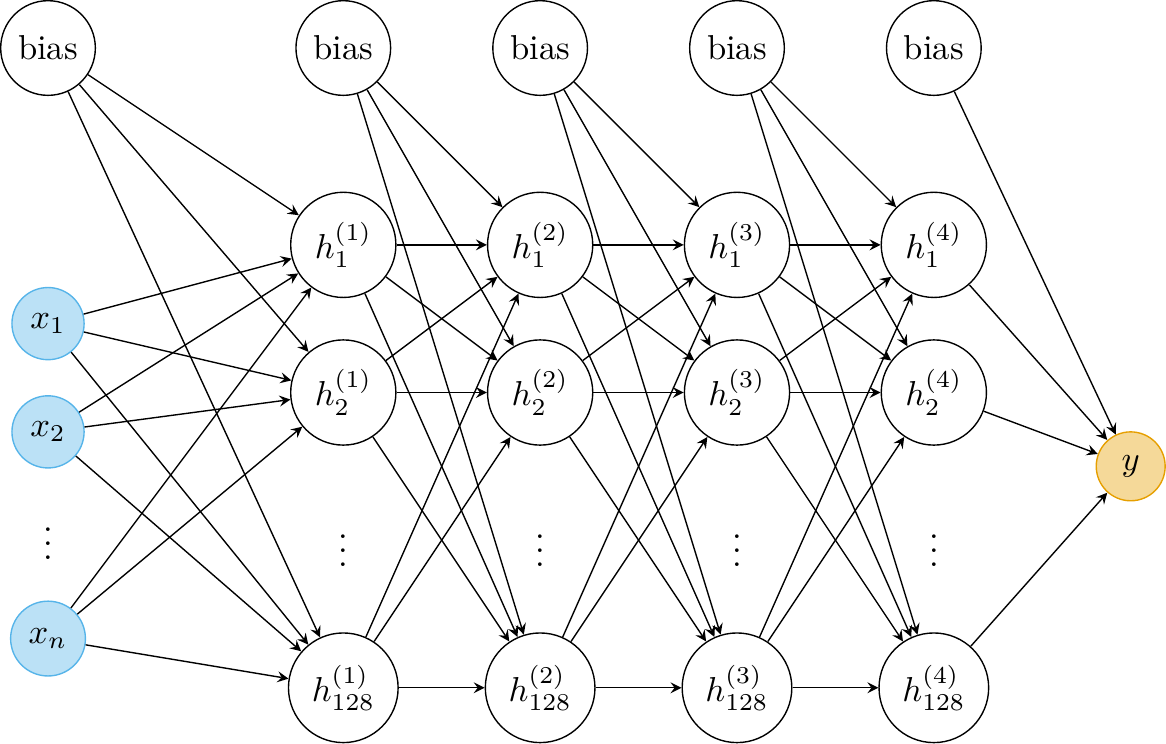}
\caption{\label{fig:figure_ann_topology} The topology of the neural network. The input variables are denoted $x_1$ to $x_n$, where the input dimension $n$ is either 2, 3, 96, 288, or 289. There are four fully-connected hidden layers with 128 nodes each. The $k$:th node of the $l$:th layer is here denoted by $h_k^l$. Rectified linear unit (ReLU) activations are used for the hidden layers. The output $y$ is just the logarithm of the permeability.}
\end{figure}

\begin{figure}[h]
\centering\includegraphics[width=80mm]{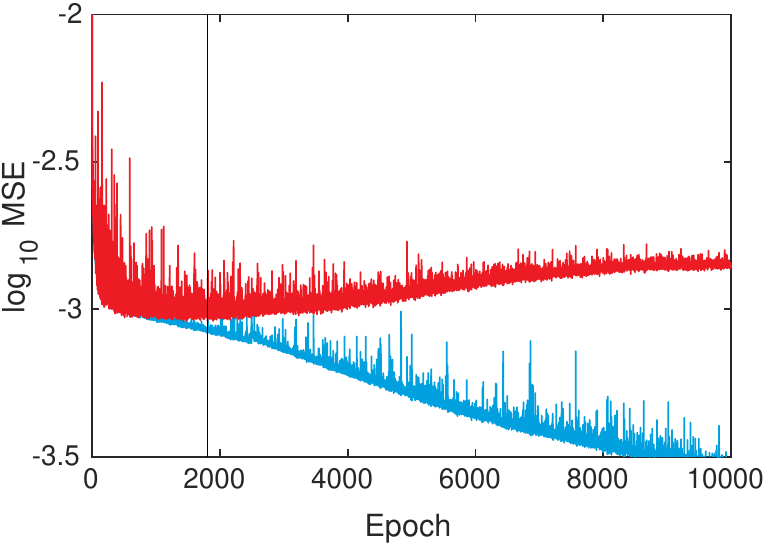}
\caption{\label{fig:figure_loss_curves} An example of a training (blue) and validation (red) loss curves for model 6$'$. This is the best run for this model, i.e., the one that yielded the minimum validation MSE, which is indicated by the vertical line (black).}
\end{figure}

We use four fully-connected hidden layers with 128 nodes each and rectified linear unit (ReLU) activations. Given that the input dimension is $n$, and the output dimension is unity, the number of weights in the network is $(n+1) \times 128 + 3 \times 129 \times 128 + 128 + 1$, i.e., there are between 50,049 and 86,785 weights to be optimized. The network is shown in Fig.\ \ref{fig:figure_ann_topology}.
Random initial weights are selected using the Glorot/Xavier uniform initializer. The networks are trained using the Adam optimizer \cite{Kingma2014} with learning rate $10^{-4}$, batch size 128, and mean squared error loss. All models are trained 100 times for 10,000 epochs using different random weight initializations, and the models with the globally minimal validation loss (MSE) are selected (hence utilizing early stopping, but performed over multiple realizations/initializations). The reason for this procedure is to minimize the impact of the random weight initializations. The models are implemented in TensorFlow 2.1.0 (\texttt{www.tensorflow.org}) \cite{Tensorflow2015}. An example of training and validation loss curves is shown in Fig.\ \ref{fig:figure_loss_curves}.
Again, the errors for training, validation, and test sets are shown in Table \ref{tab:results_ann}.
\begin{table}[h]
\centering
\begin{tabular}{p{0.75cm} p{1.75cm} p{1.75cm} p{1.75cm} p{1.75cm}}
\hline
            & RMSE          &           &           & MAPE (\%)     \\
No.         & Train         & Val       & Test      & Test          \\
\hline
\multicolumn{5}{l}{\textit{Unscaled models}}                        \\
1           & 0.029         & 0.030     & 0.041     & 6.374         \\
2           & 0.317         & 0.322     & 0.331     & 76.975        \\
3           & 0.032         & 0.035     & 0.037     & 5.479         \\
4           & 0.031         & 0.032     & 0.033     & 4.705         \\
5           & 0.057         & 0.066     & 0.066     & 10.760        \\
6           & 0.028         & 0.030     & 0.032     & 4.431         \\
7           & 0.021         & 0.023     & 0.025     & 3.679         \\
\multicolumn{5}{l}{\textit{Rescaled models}}                        \\
1$'$        & 0.045         & 0.045     & 0.046     & 7.857         \\
2$'$        & 0.326         & 0.330     & 0.339     & 82.826        \\
3$'$        & 0.032         & 0.037     & 0.039     & 6.101         \\
4$'$        & 0.031         & 0.034     & 0.036     & 5.179         \\
5$'$        & 0.056         & 0.067     & 0.069     & 11.080        \\
6$'$        & 0.029         & 0.032     & 0.036     & 5.176         \\
7$'$        & 0.020         & 0.026     & 0.028     & 4.133         \\
\hline
\end{tabular}
\caption{RMSE for the training, validation, and test sets and MAPE for the test set for the ANN models.}
\label{tab:results_ann}
\end{table}
In Fig.\ \ref{fig:figure_prediction_ann}, the predicted values vs the true (simulated) values are shown. Indeed, the neural networks based regressions perform noticeably better than the linear counterpart. Again, we see that the combination of correlation functions and geodesic tortuosity gives the best performance, achieving an impressive MAPE that is less than $4\%$. We also notice that all correlation function based models (except for $F_\mathrm{ss}$ and $F$ for aforementioned reasons) perform very well, and all better than the Kozeny-Carman-like model.

\begin{figure}[h]
\centering\includegraphics[width=80mm]{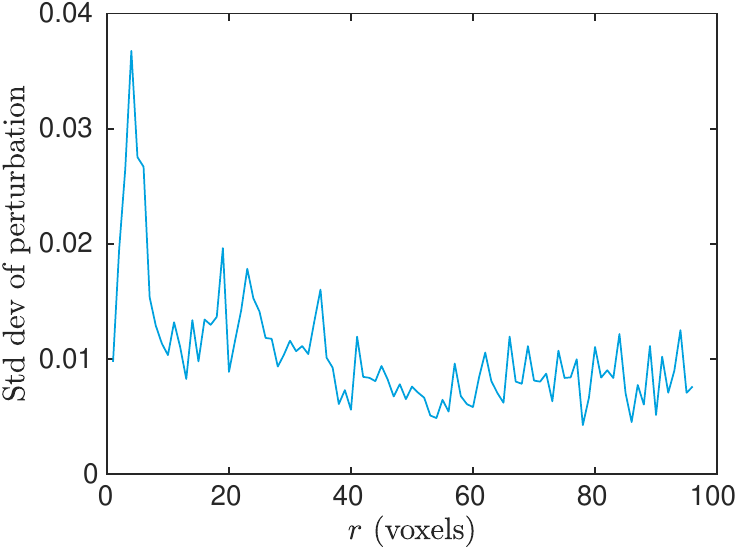}
\caption{\label{fig:figure_ann_sensitivity_model_4} The standard deviation of the perturbation in the output of the neural network for model 4, as a function of the $r$ value where the perturbation in the input is introduced.}
\end{figure}

To gain some understanding of the neural network and how the prediction is performed, we perform for the case of model 4 an analysis of the network's sensitivity with respect to perturbations in the input. Specifically, for the test set, we add random Gaussian noise to $F_\mathrm{vv}(r)$ for one $r$ value at a time. The perturbation in the output is quantified by the standard deviation of the difference between the original prediction and the perturbed prediction. In Fig.\ \ref{fig:figure_ann_sensitivity_model_4}, we show the results for $\sigma = 0.02$ (it turns out that for a broad range of perturbations, $0.0001 \le \sigma \le 0.1$, the result changes only by a constant scaling).
We see a rough resemblance to Fig.\ \ref{fig:figure_coeff_fun_model_4} in the sense that large magnitudes are mostly found for small $r$, indicating that to some extent the models are using the same information in the correlation function.

\section*{Conclusions and discussion}
We have studied data-driven structure-property relationships between fluid permeabilities and a variety of microstructural descriptors in a large data set of 30,000 virtual, porous microstructures of different types. The data set includes both granular and continuous solid phases, and is the largest one ever generated for the study of transport properties to our knowledge. To characterize the pore space geometry, we computed one-point correlation functions (porosity, specific surface), two-point surface-surface, surface-void, and void-void correlation functions, and geodesic tortuosity. Different combinations of these descriptors were used as input for different statistical learning methods, including linear regression with linear and quadratic terms, as well as deep neural networks. We find that the performance improves as the regression models become more complex, suggesting the complex relationship between the structural descriptors and the physical properties. Sufficiently large neural networks are able to fully capture the information content of the descriptors and reveal their utilities. With higher-order descriptors, we obtain significant improvements of performance when compared to a Kozeny-Carman regression with only lowest-order descriptors (porosity and specific surface). We found that combining all three two-point correlation functions and tortuosity provides the best prediction of permeability. The void-void correlation function was found to be the most informative individual descriptor. Also, the combination of porosity, specific surface, and geodesic tortuosity provides comparable predictive performance, in spite of its simplicity. Indeed, this shows that the greater information content contained in higher-order correlation functions are extremely useful for predicting physical properties of complex materials. Moreover, our work demonstrates that advanced machine learning methods can be very useful in establishing structure-property relationships.

An interesting observation is that in general the rescaling of permeabilities seem to improve the performance of the simple linear model, as seen from Table \ref{tab:results_reg_lin} that all models except the Kozeny-Carman-like one outperform their unscaled counterparts. However, as we add quadratic terms and the predictive model becomes more complex, the advantage of rescaling does not hold any more. Finally, for the highly nonlinear neural networks, the relative performance is completely inverted. This may suggest that the relation between the permeability and the specific surface cannot be captured by a simple rescaling, and doing so may reduce the information content contained in the original data. These observed effects of rescaling may lead to some guidelines on developing physics-aware machine learning models for other physical properties as well.

As a final remark, we emphasize that by incorporating microstructures with different length scales in our data set we make our models very robust and can be applied to real-world data. Since the permeability has the dimension $L^2$, we can easily obtain the permeability of a sample with a different length scale but the same microstructure. However, there is no universally applicable characteristic microstructural length to scale the permeability that enables a comparison of permeabilities for different microstructures. For example, for sphere packings the natural choice can be the radii of particles, but for continuous structures we need to resort to other quantities \cite{torquato2020predicting}. Thus, by training our models on samples with varying length scales, we circumvent this problem by only requiring a rescaling to the right order of magnitude. Finally, we make the data and code used publicly available to facilitate further development of permeability prediction methods \cite{Zenodo2020}.

\section*{Acknowledgements}
M.R. acknowledges the financial support of the Swedish Research Council (grant number 2016-03809) and the Swedish Research Council for Sustainable Development (grant number 2019-01295). Z. M. and S. T. acknowledge the support of the Air Force Office of Scientific Research Program on Mechanics of Multifunctional Materials and Microsystems under award No. FA9550-18-1-0514. The computations were in part performed on resources at Chalmers Centre for Computational Science and Engineering (C3SE) provided by the Swedish National Infrastructure for Computing (SNIC). A GPU used for part of this research was donated by the NVIDIA Corporation. Tobias Geb\"{a}ck is acknowledged for help concerning the lattice Boltzmann computations. Victor W{\aa}hlstrand Sk\"{a}rstr\"{o}m is acknowledged for producing Fig.\ \ref{fig:figure_ann_topology}.

\section*{Author contributions}
M.R., Z.M., and S.T. conceived the project and designed the study, and S.T. supervised the work. M.R. and Z.M. performed the computations and analysis. M.R., Z.M, and S.T. wrote and prepared the manuscript. M.R. and Z.M. contributed equally to this work.

\section*{Competing interests}
The authors declare no competing interests.

\section*{Data availability}
The data, i.e.\ microstructural descriptors and computed permeabilities, together with the trained models and the code used are publicly available via a repository \cite{Zenodo2020}.

\providecommand{\noopsort}[1]{}\providecommand{\singleletter}[1]{#1}%

%\bibliography{paper}

\begin{thebibliography}{67}
\providecommand{\natexlab}[1]{#1}
\providecommand{\url}[1]{\texttt{#1}}
\expandafter\ifx\csname urlstyle\endcsname\relax
  \providecommand{\doi}[1]{doi: #1}\else
  \providecommand{\doi}{doi: \begingroup \urlstyle{rm}\Url}\fi

\bibitem[Torquato(2013)]{Torquato2013}
Torquato, S.
\newblock \emph{Random heterogeneous materials: microstructure and macroscopic
  properties}.
\newblock Springer, 2013.

\bibitem[Vasseur et~al.(2020)Vasseur, Wadsworth, and
  Dingwell]{vasseur2020permeability}
Vasseur, J., Wadsworth, F.~B., and Dingwell, D.~B.
\newblock Permeability of polydisperse magma foam.
\newblock \emph{Geology}, 2020.

\bibitem[Silvestre et~al.(2011)Silvestre, Duraccio, and Cimmino]{Silvestre2011}
Silvestre, C., Duraccio, D., and Cimmino, S.
\newblock Food packaging based on polymer nanomaterials.
\newblock \emph{Prog. Polym. Sci.}, 36:\penalty0 1766--1782, 2011.

\bibitem[Slater and Cooper(2015)]{Slater2015}
Slater, A. and Cooper, A.
\newblock Function-led design of new porous materials.
\newblock \emph{Science}, 348:\penalty0 aaa8075, 2015.

\bibitem[Stamenkovic et~al.(2017)Stamenkovic, Strmcnik, Lopes, and
  Markovic]{Stamenkovic2017}
Stamenkovic, V., Strmcnik, D., Lopes, P., and Markovic, N.
\newblock Energy and fuels from electrochemical interfaces.
\newblock \emph{Nat. Mater.}, 16:\penalty0 57--69, 2017.

\bibitem[van Langenhove(2007)]{Langenhove2007}
van Langenhove, L.
\newblock \emph{Smart textiles for medicine and healthcare: materials, systems
  and applications}.
\newblock Elsevier, 2007.

\bibitem[Marucci et~al.(2013)Marucci, Andersson, Hj{\"a}rtstam, Stevenson,
  Baderstedt, Stading, Larsson, and von Corswant]{Marucci2013}
Marucci, M., Andersson, H., Hj{\"a}rtstam, J., Stevenson, G., Baderstedt, J.,
  Stading, M., Larsson, A., and von Corswant, C.
\newblock New insights on how to adjust the release profile from coated pellets
  by varying the molecular weight of ethyl cellulose in the coating film.
\newblock \emph{Int. J. Pharm.}, 458:\penalty0 218--223, 2013.

\bibitem[Milton and Sawicki(2003)]{milton2003theory}
Milton, G. and Sawicki, A.
\newblock Theory of composites. cambridge monographs on applied and
  computational mathematics.
\newblock \emph{Appl. Mech. Rev.}, 56:\penalty0 B27--B28, 2003.

\bibitem[Sahimi(2011)]{sahimi2011flow}
Sahimi, M.
\newblock \emph{Flow and transport in porous media and fractured rock: from
  classical methods to modern approaches}.
\newblock John Wiley \& Sons, 2011.

\bibitem[Blunt et~al.(2013)Blunt, Bijeljic, Dong, Gharbi, Iglauer, Mostaghimi,
  Paluszny, and Pentland]{blunt2013pore}
Blunt, M.~J., Bijeljic, B., Dong, H., Gharbi, O., Iglauer, S., Mostaghimi, P.,
  Paluszny, A., and Pentland, C.
\newblock Pore-scale imaging and modelling.
\newblock \emph{Adv. Water Resour.}, 51:\penalty0 197--216, 2013.

\bibitem[Lee et~al.(2017)Lee, Chang, Han, Kim, and Kim]{Lee2017}
Lee, S.-H., Chang, W.-S., Han, S.-M., Kim, D.-H., and Kim, J.-K.
\newblock Synchrotron x-ray nanotomography and three-dimensional nanoscale
  imaging analysis of pore structure-function in nanoporous polymeric
  membranes.
\newblock \emph{J. Membr. Sci.}, 535:\penalty0 28--34, 2017.

\bibitem[Gunda et~al.(2011)Gunda, Choi, Berson, Kenney, Karan, Pharoah, and
  Mitra]{Gunda2011}
Gunda, N., Choi, H.-W., Berson, A., Kenney, B., Karan, K., Pharoah, J., and
  Mitra, S.
\newblock Focused ion beam-scanning electron microscopy on solid-oxide
  fuel-cell electrode: Image analysis and computing effective transport
  properties.
\newblock \emph{J. Power Sources}, 196:\penalty0 3592--3603, 2011.

\bibitem[Kozeny(1927)]{Kozeny1927}
Kozeny, J.
\newblock {\"U}ber kapillare leitung des wassers im boden:(aufstieg,
  versickerung und anwendung auf die bew{\"a}sserung).
\newblock \emph{Sitz. Ber. Akad. Wiss, Wien, Math. Nat.}, 136:\penalty0
  271--306, 1927.

\bibitem[Carman(1937)]{Carman1937}
Carman, P.
\newblock Fluid flow through granular beds.
\newblock \emph{Trans. Inst. Chem. Eng.}, 15:\penalty0 150--166, 1937.

\bibitem[Kaviany(2012)]{kaviany2012principles}
Kaviany, M.
\newblock \emph{Principles of heat transfer in porous media}.
\newblock Springer Science \& Business Media, 2012.

\bibitem[Xu and Yu(2008)]{Xu2008}
Xu, P. and Yu, B.
\newblock Developing a new form of permeability and kozeny--carman constant for
  homogeneous porous media by means of fractal geometry.
\newblock \emph{Adv. Water Resour.}, 31:\penalty0 74--81, 2008.

\bibitem[Mauret and Renaud(1997)]{mauret1997transport}
Mauret, E. and Renaud, M.
\newblock Transport phenomena in multi-particle systems - i. limits of
  applicability of capillary model in high voidage beds-application to fixed
  beds of fibers and fluidized beds of spheres.
\newblock \emph{Chem. Eng. Sci.}, 52:\penalty0 1807--1817, 1997.

\bibitem[Mota et~al.(2001)Mota, Teixeira, Bowen, and Yelshin]{mota2001binary}
Mota, M., Teixeira, J., Bowen, W., and Yelshin, A.
\newblock Binary spherical particle mixed beds: porosity and permeability
  relationship measurement.
\newblock \emph{Trans. Filtr. Soc.}, 1:\penalty0 101--106, 2001.

\bibitem[Plessis and Masliyah(1991)]{du1991flow}
Plessis, J.~D. and Masliyah, J.
\newblock Flow through isotropic granular porous media.
\newblock \emph{Transport Porous Med.}, 6:\penalty0 207--221, 1991.

\bibitem[Ahmadi et~al.(2011)Ahmadi, Mohammadi, and
  Hayati]{ahmadi2011analytical}
Ahmadi, M., Mohammadi, S., and Hayati, A.
\newblock Analytical derivation of tortuosity and permeability of monosized
  spheres: A volume averaging approach.
\newblock \emph{Phys. Rev. E}, 83:\penalty0 026312, 2011.

\bibitem[Jiao et~al.(2009)Jiao, Stillinger, and Torquato]{jiao2009superior}
Jiao, Y., Stillinger, F., and Torquato, S.
\newblock A superior descriptor of random textures and its predictive capacity.
\newblock \emph{Proc. Natl. Acad. Sci.}, 106:\penalty0 17634--17639, 2009.

\bibitem[Gommes et~al.(2012)Gommes, Jiao, and
  Torquato]{gommes2012microstructural}
Gommes, C., Jiao, Y., and Torquato, S.
\newblock Microstructural degeneracy associated with a two-point correlation
  function and its information content.
\newblock \emph{Phys. Rev. E}, 85:\penalty0 051140, 2012.

\bibitem[Torquato(1991)]{Torquato1991}
Torquato, S.
\newblock Random heterogeneous media: microstructure and improved bounds on
  effective properties.
\newblock \emph{Appl. Mech. Rev.}, 44:\penalty0 37--76, 1991.

\bibitem[Jiao and Torquato(2012)]{Jiao2012}
Jiao, Y. and Torquato, S.
\newblock Quantitative characterization of the microstructure and transport
  properties of biopolymer networks.
\newblock \emph{Phys. Biol.}, 9:\penalty0 036009, 2012.

\bibitem[Prager(1961)]{Prager1961}
Prager, S.
\newblock Viscous flow through porous media.
\newblock \emph{Phys. Fluids}, 4:\penalty0 1477--1482, 1961.

\bibitem[Weissberg and Prager(1962)]{Weissberg1962}
Weissberg, H. and Prager, S.
\newblock Viscous flow through porous media. ii. approximate three-point
  correlation function.
\newblock \emph{Phys. Fluids}, 5:\penalty0 1390--1392, 1962.

\bibitem[Weissberg and Prager(1970)]{Weissberg1970}
Weissberg, H. and Prager, S.
\newblock Viscous flow through porous media. iii. upper bounds on the
  permeability for a simple random geometry.
\newblock \emph{Phys. Fluids}, 13:\penalty0 2958--2965, 1970.

\bibitem[Berryman and Milton(1985)]{Berryman1985a}
Berryman, J. and Milton, G.
\newblock Normalization constraint for variational bounds on fluid
  permeability.
\newblock \emph{J. Chem. Phys.}, 83:\penalty0 754--760, 1985.

\bibitem[Berryman(1985)]{Berryman1985b}
Berryman, J.
\newblock Bounds on fluid permeability for viscous flow through porous media.
\newblock \emph{J. Chem. Phys.}, 82:\penalty0 1459--1467, 1985.

\bibitem[Rubinstein and Torquato(1989)]{Rubinstein1989}
Rubinstein, J. and Torquato, S.
\newblock Flow in random porous media: mathematical formulation, variational
  principles, and rigorous bounds.
\newblock \emph{J. Fluid Mech.}, 206:\penalty0 25--46, 1989.

\bibitem[Liasneuski et~al.(2014)Liasneuski, Hlushkou, Khirevich, H{\"o}ltzel,
  Tallarek, and Torquato]{Liasneuski2014}
Liasneuski, H., Hlushkou, D., Khirevich, S., H{\"o}ltzel, A., Tallarek, U., and
  Torquato, S.
\newblock Impact of microstructure on the effective diffusivity in random
  packings of hard spheres.
\newblock \emph{J. Appl. Phys.}, 116:\penalty0 034904, 2014.

\bibitem[Hlushkou et~al.(2015)Hlushkou, Liasneuski, Tallarek, and
  Torquato]{Hlushkou2015}
Hlushkou, D., Liasneuski, H., Tallarek, U., and Torquato, S.
\newblock Effective diffusion coefficients in random packings of polydisperse
  hard spheres from two-point and three-point correlation functions.
\newblock \emph{J. Appl. Phys.}, 118:\penalty0 124901, 2015.

\bibitem[Zachary and Torquato(2011)]{Zachary2011}
Zachary, C. and Torquato, S.
\newblock Improved reconstructions of random media using dilation and erosion
  processes.
\newblock \emph{Phys. Rev. E}, 84:\penalty0 056102, 2011.

\bibitem[Guo et~al.(2014)Guo, Chawla, Jing, Torquato, and Jiao]{Guo2014}
Guo, E.-Y., Chawla, N., Jing, T., Torquato, S., and Jiao, Y.
\newblock Accurate modeling and reconstruction of three-dimensional percolating
  filamentary microstructures from two-dimensional micrographs via
  dilation-erosion method.
\newblock \emph{Mater. Charact.}, 89:\penalty0 33--42, 2014.

\bibitem[Katz and Thompson(1986)]{katz1986quantitative}
Katz, A. and Thompson, A.
\newblock Quantitative prediction of permeability in porous rock.
\newblock \emph{Phys. Rev. B}, 34:\penalty0 8179, 1986.

\bibitem[Torquato(2020)]{torquato2020predicting}
Torquato, S.
\newblock Predicting transport characteristics of hyperuniform porous media via
  rigorous microstructure-property relations.
\newblock \emph{Adv. Water Resour.}, 140:\penalty0 103565, 2020.

\bibitem[Avellaneda and Torquato(1991)]{avellaneda1991rigorous}
Avellaneda, M. and Torquato, S.
\newblock Rigorous link between fluid permeability, electrical conductivity,
  and relaxation times for transport in porous media.
\newblock \emph{Phys. Fluids A: Fluid Dynamics}, 3:\penalty0 2529--2540, 1991.

\bibitem[Ghanbarian et~al.(2013)Ghanbarian, Hunt, Ewing, and
  Sahimi]{Ghanbarian2013}
Ghanbarian, B., Hunt, A., Ewing, R., and Sahimi, M.
\newblock Tortuosity in porous media: {A} critical review.
\newblock \emph{Soil Sci. Soc. Am. J.}, 77:\penalty0 1461--1477, 2013.

\bibitem[van~der Linden et~al.(2016)van~der Linden, Narsilio, and
  Tordesillas]{Linden2016}
van~der Linden, J., Narsilio, G., and Tordesillas, A.
\newblock Machine learning framework for analysis of transport through complex
  networks in porous, granular media: a focus on permeability.
\newblock \emph{Phys. Rev. E}, 94:\penalty0 022904, 2016.

\bibitem[Stenzel et~al.(2016)Stenzel, Pecho, Holzer, Neumann, and
  Schmidt]{Stenzel2016}
Stenzel, O., Pecho, O., Holzer, L., Neumann, M., and Schmidt, V.
\newblock Predicting effective conductivities based on geometric microstructure
  characteristics.
\newblock \emph{AIChE J.}, 62:\penalty0 1834--1843, 2016.

\bibitem[Neumann et~al.(2019)Neumann, Stenzel, Willot, Holzer, and
  Schmidt]{Neumann2019}
Neumann, M., Stenzel, O., Willot, F., Holzer, L., and Schmidt, V.
\newblock Quantifying the influence of microstructure on effective conductivity
  and permeability: virtual materials testing.
\newblock \emph{Int. J. Solids Struct.}, 2019.
\newblock \doi{10.1016/j.ijsolstr.2019.03.028}.

\bibitem[Barman et~al.(2019)Barman, Rootz\'{e}n, and Bolin]{Barman2019}
Barman, S., Rootz\'{e}n, H., and Bolin, D.
\newblock Prediction of diffusive transport through polymer films from
  characteristics of the pore geometry.
\newblock \emph{AIChE J.}, 65:\penalty0 446--457, 2019.

\bibitem[Kondo et~al.(2017)Kondo, Yamakawa, Masuoka, Tajima, and
  Asahi]{Kondo2017}
Kondo, R., Yamakawa, S., Masuoka, Y., Tajima, S., and Asahi, R.
\newblock Microstructure recognition using convolutional neural networks for
  prediction of ionic conductivity in ceramics.
\newblock \emph{Acta Mater.}, 141:\penalty0 29--38, 2017.

\bibitem[Wu et~al.(2018)Wu, Yin, and Xiao]{Wu2018}
Wu, J., Yin, X., and Xiao, H.
\newblock Seeing permeability from images: fast prediction with convolutional
  neural networks.
\newblock \emph{Sci. Bull.}, 63:\penalty0 1215--1222, 2018.

\bibitem[Sudakov et~al.(2019)Sudakov, Burnaev, and Koroteev]{Sudakov2019}
Sudakov, O., Burnaev, E., and Koroteev, D.
\newblock Driving digital rock towards machine learning: predicting
  permeability with gradient boosting and deep neural networks.
\newblock \emph{Comput. Geosci.}, 127:\penalty0 91--98, 2019.

\bibitem[Stenzel et~al.(2017)Stenzel, Pecho, Holzer, Neumann, and
  Schmidt]{Stenzel2017}
Stenzel, O., Pecho, O., Holzer, L., Neumann, M., and Schmidt, V.
\newblock Big data for microstructure-property relationships: {A} case study of
  predicting effective conductivities.
\newblock \emph{AIChE J.}, 63:\penalty0 4224--4232, 2017.

\bibitem[Kamrava et~al.(2020)Kamrava, Tahmasebi, and Sahimi]{Kamrava2020}
Kamrava, S., Tahmasebi, P., and Sahimi, M.
\newblock Linking morphology of porous media to their macroscopic permeability
  by deep learning.
\newblock \emph{Transport Porous Med.}, 131:\penalty0 427--448, 2020.

\bibitem[Wu et~al.(2019)Wu, Fang, Kang, Tao, and Qiao]{Wu2019}
Wu, H., Fang, W.-Z., Kang, Q., Tao, W.-Q., and Qiao, R.
\newblock Predicting effective diffusivity of porous media from images by deep
  learning.
\newblock \emph{Sci. Rep.}, 9:\penalty0 20387, 2019.

\bibitem[Lubbers et~al.(2017)Lubbers, Lookman, and Barros]{Lubbers2017}
Lubbers, N., Lookman, T., and Barros, K.
\newblock Inferring low-dimensional microstructure representations using
  convolutional neural networks.
\newblock \emph{Phys. Rev. E}, 96:\penalty0 052111, 2017.

\bibitem[R\"oding et~al.(2020)R\"oding, Ma, and Torquato]{Zenodo2020}
R\"oding, M., Ma, Z., and Torquato, S.
\newblock \emph{zenodo} \url{http://dx.doi.org/10.5281/zenodo.3752765}, 2020.

\bibitem[Pecho et~al.(2015)Pecho, Stenzel, Iwanschitz, Gasser, Neumann,
  Schmidt, Prestat, Hocker, Flatt, and Holzer]{Pecho2015}
Pecho, O., Stenzel, O., Iwanschitz, B., Gasser, P., Neumann, M., Schmidt, V.,
  Prestat, M., Hocker, T., Flatt, R., and Holzer, L.
\newblock {3D} microstructure effects in {Ni-YSZ} anodes: {P}rediction of
  effective transport properties and optimization of redox stability.
\newblock \emph{Materials}, 8:\penalty0 5554--5585, 2015.

\bibitem[Ma and Torquato(2018)]{Ma2018}
Ma, Z. and Torquato, S.
\newblock Precise algorithms to compute surface correlation functions of
  two-phase heterogeneous media and their applications.
\newblock \emph{Phys. Rev. E}, 98:\penalty0 013307, 2018.

\bibitem[Scholz et~al.(2015)Scholz, Wirner, Klatt, Hirneise, Schr{\"o}der-Turk,
  Mecke, and Bechinger]{scholz2015direct}
Scholz, C., Wirner, F., Klatt, M., Hirneise, D., Schr{\"o}der-Turk, G., Mecke,
  K., and Bechinger, C.
\newblock Direct relations between morphology and transport in boolean models.
\newblock \emph{Phys. Rev. E}, 92:\penalty0 043023, 2015.

\bibitem[Howard et~al.(2020)Howard, Lequieu, Delaney, Ganesan, Fredrickson, and
  Truskett]{howard2020connecting}
Howard, M., Lequieu, J., Delaney, K., Ganesan, V., Fredrickson, G., and
  Truskett, T.
\newblock Connecting solute diffusion to morphology in triblock copolymer
  membranes.
\newblock \emph{Macromolecules}, 2020.

\bibitem[Lang and Potthoff(2011)]{Lang2011}
Lang, A. and Potthoff, J.
\newblock Fast simulation of gaussian random fields.
\newblock \emph{Monte Carlo Methods Appl.}, 17:\penalty0 195--214, 2011.

\bibitem[Mat{\'e}rn(1986)]{Matern1986}
Mat{\'e}rn, B.
\newblock \emph{Spatial variation}, volume~36.
\newblock Springer Science \& Business Media, 1986.

\bibitem[Geb{\"a}ck and Heintz(2014)]{Geback2014}
Geb{\"a}ck, T. and Heintz, A.
\newblock A lattice boltzmann method for the advection-diffusion equation with
  neumann boundary conditions.
\newblock \emph{Commun. Comput. Phys.}, 15:\penalty0 487--505, 2014.

\bibitem[Geb\"ack et~al.(2015)Geb\"ack, Marucci, Boissier, Arnehed, and
  Heintz]{Geback2015}
Geb\"ack, T., Marucci, M., Boissier, C., Arnehed, J., and Heintz, A.
\newblock Investigation of the effect of the tortuous pore structure on water
  diffusion through a polymer film using lattice boltzmann simulations.
\newblock \emph{J. Phys. Chem. B}, 119:\penalty0 5220--5227, 2015.

\bibitem[Perram and Wertheim(1985)]{Perram1985}
Perram, J. and Wertheim, M.
\newblock Statistical mechanics of hard ellipsoids. {I}. {O}verlap algorithm
  and the contact function.
\newblock \emph{J. Comput. Phys.}, 58:\penalty0 409--416, 1985.

\bibitem[Bezanson et~al.(2017)Bezanson, Edelman, Karpinski, and
  Shah]{Bezanson2017}
Bezanson, J., Edelman, A., Karpinski, S., and Shah, V.
\newblock Julia: A fresh approach to numerical computing.
\newblock \emph{SIAM Rev.}, 59:\penalty0 65--98, 2017.

\bibitem[Ginzburg et~al.(2008)Ginzburg, Verhaeghe, and
  d?Humieres]{Ginzburg2008}
Ginzburg, I., Verhaeghe, F., and d?Humieres, D.
\newblock Study of simple hydrodynamic solutions with the two-relaxation-times
  lattice {B}oltzmann scheme.
\newblock \emph{Commun. Comput. Phys.}, 3:\penalty0 519--581, 2008.

\bibitem[Zou and He(1997)]{Zou1997}
Zou, Q. and He, X.
\newblock On pressure and velocity boundary conditions for the lattice
  {B}oltzmann {BGK} model.
\newblock \emph{Phys. Fluids}, 9:\penalty0 1591--1598, 1997.

\bibitem[Ma and Torquato(2017)]{ma2017random}
Ma, Z. and Torquato, S.
\newblock Random scalar fields and hyperuniformity.
\newblock \emph{J. Appl. Phys.}, 121:\penalty0 244904, 2017.

\bibitem[R{\"o}ding et~al.(2017)R{\"o}ding, Svensson, and
  Lor{\'e}n]{Roding2017}
R{\"o}ding, M., Svensson, P., and Lor{\'e}n, N.
\newblock Functional regression-based fluid permeability prediction in
  monodisperse sphere packings from isotropic two-point correlation functions.
\newblock \emph{Comput. Mater. Sci.}, 134:\penalty0 126--131, 2017.

\bibitem[Tibshirani(1996)]{Tibshirani1996}
Tibshirani, R.
\newblock Regression shrinkage and selection via the lasso.
\newblock \emph{J. R. Stat. Soc. B}, 58:\penalty0 267--288, 1996.

\bibitem[Kingma and Ba(2014)]{Kingma2014}
Kingma, D. and Ba, J.
\newblock Adam: {A} method for stochastic optimization.
\newblock \emph{arXiv preprint arXiv:1412.6980}, 2014.

\bibitem[Abadi et~al.(2015)Abadi, Agarwal, Barham, Brevdo, Chen, Citro,
  Corrado, Davis, Dean, Devin, Ghemawat, Goodfellow, Harp, Irving, Isard, Jia,
  Jozefowicz, Kaiser, Kudlur, Levenberg, Man\'{e}, Monga, Moore, Murray, Olah,
  Schuster, Shlens, Steiner, Sutskever, Talwar, Tucker, Vanhoucke, Vasudevan,
  Vi\'{e}gas, Vinyals, Warden, Wattenberg, Wicke, Yu, and
  Zheng]{Tensorflow2015}
Abadi, M., Agarwal, A., Barham, P., Brevdo, E., Chen, Z., Citro, C., Corrado,
  G.~S., Davis, A., Dean, J., Devin, M., Ghemawat, S., Goodfellow, I., Harp,
  A., Irving, G., Isard, M., Jia, Y., Jozefowicz, R., Kaiser, L., Kudlur, M.,
  Levenberg, J., Man\'{e}, D., Monga, R., Moore, S., Murray, D., Olah, C.,
  Schuster, M., Shlens, J., Steiner, B., Sutskever, I., Talwar, K., Tucker, P.,
  Vanhoucke, V., Vasudevan, V., Vi\'{e}gas, F., Vinyals, O., Warden, P.,
  Wattenberg, M., Wicke, M., Yu, Y., and Zheng, X.
\newblock {TensorFlow}: Large-scale machine learning on heterogeneous systems,
  2015.
\newblock URL \url{https://www.tensorflow.org/}.
\newblock Software available from tensorflow.org.

\end{thebibliography}
%\bibliographystyle{myunsrtnat}
\end{document}